\documentclass[showpacs,aps,prd,floatfix,amsmath,amssymb]{revtex4}
\usepackage{graphicx}
\usepackage{epstopdf}
\usepackage[utf8]{inputenc}
\usepackage{amsmath}
\usepackage{array}
\usepackage{slashed}	
\usepackage[makeroom]{cancel}

\begin{document}
\title{Decay $\phi\to Z \gamma\gamma$ ($\phi=h, H,A$)  in the minimal supersymmetric standard model}
\author{R. S\'anchez-V\'elez}
\author{G. Tavares-Velasco}
\email[Corresponding author: ]{gtv@fcfm.buap.mx}
\affiliation{Facultad de
Ciencias F\'\i sico Matem\'aticas, Benem\'erita Universidad
Aut\'onoma de Puebla, Apartado Postal 1152, Puebla, Pue., M\'
exico}

\begin{abstract}
The  decays  of the CP-even ($h,\,H$) and the CP-odd ($A$) Higgs bosons $h,\,H,\,A\to Z\gamma\gamma$  are calculated in the context of the minimal supersymmetric standard model (MSSM), where they are induced at the one-loop level via box and reducible Feynman diagrams.  For the numerical evaluation of the  branching ratios  we employ the decoupling limit and consider values for the MSSM parameters still consistent  with the current experimental data. We consider a particular scenario for the radiative corrections to the Higgs boson masses, dubbed hMSSM,  which leads to only two free parameters, namely $m_A$ and $\tan\beta$. We found that the branching ratio of the $CP$-odd Higgs boson decay $A\to Z\gamma\gamma$ can reach values up to $10^{-8}$ for $m_A=750$ GeV, which is three order of magnitude  smaller than that of the $A\to\gamma\gamma$ channel. On the other hand, the branching ratio of the $H\to Z\gamma\gamma$ decay is of the order of $10^{-7}$ for large values of $m_H$ and small values of $\tan\beta$, which is comparable with the branching ratio of the $H\to Z\gamma$ channel. As far as the branching ratio of the  SM-like Higgs boson decay $h\to Z\gamma\gamma$ is concerned, it is considerably suppressed, of the order of $10^{-11}$.
\end{abstract}
\maketitle

\section{Introduction}

The discovery of a Higgs boson with a 125 GeV mass in the Run 1 of the Large Hadron Collider (LHC) by the ATLAS and CMS Collaborations \cite{Aad:2012tfa,Chatrchyan:2012xdj} is  a milestone in particle physics and the standard model (SM),  which only predicts one of such particles. Nonetheless, the  search for additional Higgs bosons predicted by theories with  extended scalar sectors is still underway. A plethora of beyond the SM theories (BSMTs)  of this class have been proposed to improve the SM but also to explore effects absent in this theory. High energy experiments have looked for signals of new physics (NP), either through the direct production of new particles or the search for evidences of anomalous contributions to the couplings of the  SM particles. However, up to now no evidences along these lines have been found, thereby shifting the bounds on the associated scales into the TeV scale. In spite of this scenario, the search for new Higgs bosons may provide signals of NP and give a hint to the path to extend the SM. Hopefully any additional scalar boson would show up at the LHC in a near future.

Among the most popular BSMTs, the minimal supersymmetric standard model (MSSM) \cite{Martin:1997ns,Gunion85nuclearphysics} stands out and has been the focus of considerable attention in the literature during the last decades. Supersymmetry (SUSY) associate bosons with fermions, predicting the existence of  fermion partners for all SM bosons  as well as scalar partners for all SM fermions. Unlike the SM scalar sector, which requires only one Higgs doublet to trigger the electroweak symmetry breaking, two Higgs doublets are  necessary in the framework of the MSSM. After the electroweak symmetry is broken,  five physical Higgs bosons emerge: two $CP$-even Higgs bosons $H$ and $h$, with  $m_H>m_h$ by convention, a $CP$-odd Higgs boson $A$, often called pseudoscalar, and a pair of charged Higgs bosons $H^\pm$. The properties of one of the neutral $CP$-even scalar bosons must match those of  the 125 GeV scalar boson detected at the LHC. Of particular importance for our work are  the superpartners of the Higgs bosons and gauge bosons (Higgsinos and gauginos), which  mix to give rise to neutralino and chargino states ${\chi}^\pm$. Extensive studies have been carried out to constraint the allowed region of the parameter space of the MSSM via low energy observables along with the data from direct searches at the LHC \cite{Carena2012,Arbey2012}. Furthermore,   the LHC data on the 125 GeV Higgs boson obtained at $\sqrt{s}=7$ and 8 GeV  are in agreement with the  SM predictions for the Higgs boson couplings to the heavier SM particles, thereby placing tight constraints on the parameter space of BSMTs. Along this line,  it has been found that LHC data gives rise to a strong tension with the naturalness within the MSSM since  the mass of the stop sector is required to be rather heavy, typically above 1 TeV.

One-loop induced Higgs boson decays into vector bosons such as $\phi\to \gamma\gamma$ and $\phi\to Z\gamma$ ($\phi=h,H,A$) have been long studied in the literature within the context of several BSMTs.  These decays are interesting since they are induced by loops carrying heavy charged particles and  are thus sensitive to the  charge and spin of new charged particles. Within the MSSM, a potential enhancement of the $\phi \to \gamma\gamma$ decay width may arise from loops carrying charged Higgs bosons, sfermions, and charginos.
This possibility was considered in \cite{Carena:2012gp}  to explain the apparent excess of the diphoton decay rate of the 125 GeV Higgs boson  reported by the ATLAS and CMS Collaborations \cite{Aad:2012tfa,Chatrchyan:2012xdj}. In the scenario with  heavy sfermions, the chargino contribution is the only viable option that can induce a significant enhancement of the $h\to \gamma\gamma$ decay rate, which  can only reach the $10 \%$ level for typical values of the SUSY parameters, but it can be considerably larger for light charginos and low $\tan\beta$ \cite{Carena:2012gp}. In particular, there is a narrow area of the MSSM parameter space where the $h \to \gamma\gamma$ rate can be enhanced up to $25\%$, whereas a less substantial increase of  around $10-20\%$ can be reached in a much wider area  \cite{Casas:2013pta}. Similar studies have been carried out for the $\phi Z\gamma\;(\phi=h,H,A)$ coupling \cite{Djouadi:1996yq}. Although no excess in the  $h \to \gamma\gamma$ decay rate was observed in the data collected at the LHC Run 2, it is possible that charginos can give an enhancement to the diphoton decay rate of  heavy scalar bosons over the contributions of non-SUSY particles.

If any additional Higgs scalar boson is discovered, it would be interesting to test its exotic decay modes as they may shed light on the underlying theory. In this work we study the one-loop induced three-body decays of  the $CP$-odd  and $CP$-even Higgs bosons $\phi\to Z\gamma\gamma$ ($\phi=h,H,A$)  in the context of the MSSM, where chargino loops can enhance significantly the decay rate as compared to  the fermion and squark contributions. These decays receive contributions from both box and reducible Feynman diagrams. Experimentally this decay channel has the advantage of a relatively low background since the final state contains two photons plus a pair of energetic back-to-back leptons. In addition, as occurs with the diphoton coupling, the $AZ\gamma\gamma$ and $H Z\gamma\gamma$ ones  are also sensitive to the charge and spin of the  particles running into the loops, such as the charginos of the MSSM. Within the SM framework the decays $h\to Z\gamma\gamma$ \cite{Abbasabadi:2004wq}  and the analogue one $h\to Zgg$ \cite{Abbasabadi2008} have been studied. More recently, the $A\to Z\gamma\gamma$ and $H\to Z\gamma\gamma$ decays were studied in the context of THDMs \cite{Sanchez-Velez:2018xdj}.   Also, the $\phi Z\gamma\gamma$ $ (\phi=h,H,A)$ coupling was analyzed via photon fusion  $\gamma\gamma \to Z\phi$  in   the context of the SM and the MSSM \cite{Gounaris:2001rk}. In our study, we consider the existing limits on the parameter space of the MSSM.  We work in a scenario for the radiative corrections to the Higgs boson masses, dubbed hMSSM (habemus MSSM? \cite{Djouadi:2013uqa}),  which leads to only two free parameters, namely $m_A$ and $\tan\beta$. Since the current limit on chargino masses is $m_{\tilde{\chi}^\pm}>103.5 $ GeV \cite{Patrignani:2016xqp}, it is worth assessing if there is a significant enhancement to the $\phi\to Z\gamma\gamma$ ($\phi=H,A$) decay rate from  chargino loops.

The organization of this paper is as follows. In section II a brief outline of the MSSM is presented, with focus on the Higgs sector as well as the chargino and squark sectors. Section III is devoted to the calculation of the decay $\phi\to Z \gamma\gamma$ ($\phi=H,A$) through the Passarino-Veltman reduction scheme. In Section IV a brief discussion on the allowed parameter space of the MSSM is presented along with the numerical analysis of the  $\phi\to Z \gamma\gamma$ ($\phi=H,A$) decay width. Finally in Section V we present the conclusions and outlook. The Feynman rules involved in our calculations as well as some lengthy formulas are presented in  Appendices A and B.

\section{Overview of the minimal supersymmetric standard model}

SUSY  predicts new scalar particles called squarks and sleptons, which are superpartners of the quarks and leptons. Also, gauge bosons have fermionic partners, called gauginos. In the MSSM  \cite{Csaki:1996ks}, which is the minimal realization of SUSY,  two complex Higgs doublets with opposite hypercharges are required to give masses to the up- and down-type fermions and ensuring anomaly cancellation \cite{Fayet:2014oua}. The superpartners of the physical Higgs bosons  are  known as  Higgsinos. Below we present an outline of the MSSM Higgs sector, focusing only on those aspects  relevant for our calculation.

\subsection{Higgs sector}

The Higgs sector of the MSSM is essentially the same as that of the Type II-THDM \cite{Branco:2011iw}, where two complex isodoublets are introduced

\begin{equation}
H_1=\left(\begin{array}{c}H_1^0 \\ H_1^-
\end{array}\right)\;\; \mbox{with}\;\; Y_{H_1}=-1\;\;,\;\; H_2=\left(\begin{array}{c}H_2^+ \\ H_2^0 \end{array}\right)\;\; \mbox{with}\;\; Y_{H_2}=+1.
\end{equation}
The $H_1$ doublet couples to the down-type quarks and the charged leptons while the $H_2$ Higgs doublet only couples to the up-type quarks. After the electroweak symmetry breaking, three of the original eight degrees of freedom of $H_1$ and $H_2$ become the longitudinal polarizations of the $W^\pm $ and $Z$ gauge bosons, which thus acquire masses, whereas the remaining degrees of freedom correspond to  the physical Higgs spectrum, which is determined as follows. The neutral components of the two Higgs fields develop vacuum expectations values (VEVs)

\begin{equation}
\langle H_1^0\rangle =\frac{v_1}{\sqrt{2}}\;,\;\; \langle
H_2^0\rangle = \frac{v_2}{\sqrt{2}}.
\end{equation}
To obtain the  physical Higgs fields, along with their masses,  the two doublet complex scalar fields $H_1$ and $H_2$ are expanded around their VEVs into their real and imaginary parts as follows

\begin{equation}
H_1=\frac{1}{\sqrt{2}}\left(\begin{array}{c}
v_1 +H_1^0+iP_1^0\\ H_1^- \end{array}\right),\;\;
H_2=\frac{1}{\sqrt{2}}\left(\begin{array}{c}
H_2^+ \\ v_2 +H_2^0+iP_2^0 \end{array}\right),
\end{equation}
where the real parts are  linear combination of the $CP$-even Higgs bosons $h$ and $H$, whereas the imaginary parts correspond to the $CP$-odd Higgs boson $A$ and the neutral Goldstone boson.
The physical $CP$-even Higgs bosons are obtained upon rotating the neutral components of the Higgs doublets by the angle $\alpha$

\begin{equation}
\label{alpharotation}
\left(\begin{array}{c}
H \\ h
\end{array}\right)=\left(\begin{array}{rc}
c_\alpha &s_\alpha\\
-s_\alpha &c_\alpha
\end{array}\right) \left(\begin{array}{c}
H_1^0 \\ H_2^0
\end{array}\right),
\end{equation}
where we introduced the following short-hand notation to be used from now on $s_{\zeta}\equiv \sin{\zeta}$ and $c_{\zeta}\equiv\cos{\zeta}$, where $\zeta$ stands for any angle.
In the case of the $CP$-odd Higgs boson and the neutral Goldstone boson $G^0$,  they are obtained after rotating the imaginary components of the Higgs doublets by the angle $\beta$

\begin{equation}
\left(\begin{array}{c} G^0 \\ A\end{array}\right)=
\left(\begin{array}{rc}
c_\beta & s_\beta\\
-s_\beta & c_\beta \end{array}\right)
\left(\begin{array}{c} P_1^0 \\ P_2^0
\end{array}\right),
\end{equation}
where $\tan\beta$ is  the ratio of VEVs, i.e. $\tan\beta=v_1/v_2$.

The mass matrix of the $CP$-even Higgs in the $H_1^0$-$H_2^0$ basis is given by
\begin{equation}\label{massmatrix}
M_S^2=m_Z^2
\begin{pmatrix}
c ^2_\beta & -c_\beta s _\beta \\
 -c _\beta s _\beta& s ^2_\beta
\end{pmatrix}+m_A^2
\begin{pmatrix}
s ^2_\beta & -c_\beta s _\beta \\
 -c _\beta s _\beta& c ^2_\beta
\end{pmatrix}+
\begin{pmatrix}
\Delta {\cal M}^2_{11} & \Delta {\cal M}^2_{12}\\
\Delta {\cal M}^2_{12} & \Delta {\cal M}^2_{22}
\end{pmatrix}
\end{equation}
where $\Delta {\cal M}^2_{ij}$ stands for the radiative corrections that depend on the SUSY scale, the stop/bottom trilinear couplings $A_{t/b}$, and the Higgsino mass $\mu$.
After the diagonalization of $M_S^2$  and the rotation \eqref{alpharotation}, one obtains the masses of the physical neutral Higgs bosons up to radiative corrections

\begin{equation}\label{mhma}
m_{h,H}^2=\frac{1}{2}\Bigl[ m_A^2+m_Z^2+\Delta {\cal M}^2_{11}+\Delta {\cal M}^2_{22}\mp \sqrt{(m_A^2+m_Z^2)^2-4\,m_A^2m_Z^2c^2_{ 2\beta}+C}\Bigr],
\end{equation}
where
\begin{equation}\label{C}
C=4\Delta{\cal M}^4_{12}+\left(\Delta{\cal M}^2_{11}
-\Delta{\cal M}^2_{22}\right)^2 -2\left(m_A^2-m_Z^2\right)\left(\Delta{\cal M}^2_{11}
-\Delta{\cal M}^2_{22}\right)c_{2\beta} -4\left(m_A^2+m_Z^2\right)
\Delta{\cal M}^2_{12}s_{2\beta}
\end{equation}
Note that  $m_h\le m_Z$ if radiative corrections are not  taken into account. In a simplified version of the model, the so called hMSMS \cite{Djouadi:2013uqa}, it is assumed that $\Delta{\cal M}_{22}$, which includes the dominant stop corrections, is the only relevant correction term, i.e. $\Delta{\cal M}_{22}\gg \Delta{\cal M}_{11},\,\Delta{\cal M}_{12} $. Thus if the lightest $CP$-even scalar boson $h$ is assumed to be the one found at the LHC,  $\Delta{\cal M}_{22}$ can be traded for the already known mass $m_h$ as follows \cite{Djouadi:2013uqa}
\begin{equation}\label{DeltaM22}
\Delta{\cal M}_{22}=\frac{m_h^2\left(m_A^2+m_Z^2-m_h^2\right)-m_A^2 m_Z^2 c^2_{2\beta}}
{m_Z^2 c_\beta^2+m_A^2 s_\beta^2-m_h^2}.
\end{equation}
Therefore $m_H$ can be written as
\begin{equation}
\label{mH}
m_H^2=\frac{\left(m_A^2+m_Z^2-m_h^2\right)\left(m_Z^2 c_\beta^2+m_A^2 s_\beta^2\right)-m_A^2 m_Z^2 c^2_{2\beta}}
{m_Z^2 c_\beta^2+m_A^2 s_\beta^2-m_h^2},
\end{equation}
and the mixing angle $\alpha$ obeys
\begin{equation}\label{alpha}
\tan\alpha=\frac{\left(m_A^2+m_Z^2\right) c_{\beta}s_\beta}
{m_Z^2 c_\beta^2+m_A^2 s_\beta^2-m_h^2}.
\end{equation}
There are thus two free parameters that  are usually chosen as $t_\beta\equiv \tan\beta$ and $m_A$. Our analysis below will be focused on the hMSSM scenario.

\subsection{Chargino Sector}\label{charsector}

The superpartners of the charged Higgs boson and the $W$ gauge boson mix to give rise to the chargino mass eigenstates $\tilde{\chi}_{1,2}^\pm$. The mass eigenvalues $m_{\tilde{\chi}_{1,2}^\pm}$, the mixing angles, and the phases are determined by the elements of the chargino mass matrix, which in the $(\tilde{W}^-, \tilde{H}^-)$ basis is given by

\begin{equation}\label{massmatrixchar}
X=\left( \begin{array}{cc} M_2 &\sqrt{2}m_W c_\beta \\
\sqrt{2}m_W s_\beta &\mu
\end{array}\right),
\end{equation}
thereby being determined by the fundamental SUSY parameters: the gaugino mass $M_2$, the Higgs mass $\mu$ and the mixing angle $t_\beta$. In several models, the lightest chargino state $\tilde{\chi}_1^+$ is expected to be one of the lightest SUSY particle and might play an important role in the experimental detection of SUSY.

The Lagrangian for the mass terms of the charginos can be written in the gaugino-Higgsino eigenbasis as

\begin{equation}
\mathcal{L}_{\chi^c \mbox{mass}} =-\Psi^T_R X \Psi_L+ \mbox{H.c},
\end{equation}
where $\Psi_L$ and $\Psi_R$ are two-components Weyl spinors

\begin{equation}
\Psi_L=\begin{pmatrix}
         \tilde{W}^+ \\
         \tilde{h}_2^+
       \end{pmatrix},\;\;
       \Psi_R=\begin{pmatrix}
         \tilde{W}^- \\
         \tilde{h}_1^-
       \end{pmatrix}.
\end{equation}

The chargino mass matrix $X$ can be diagonalized by rotating the  wino and Higgsino fields via  $2\times 2$  unitary matrices, namely, $\chi_L=V\Psi_L$ and $\chi_R=U\Psi_R$, with  $\chi_{L,R}$ being the chargino mass eigenstates. This leads to

\begin{equation}\label{unit}
M_{\tilde{\chi^+}}=UXV^T=\begin{pmatrix}
m_{\tilde{\chi}^+_1}& 0\\
0 & m_{\tilde{\chi}^+_2}
 \end{pmatrix},
\end{equation}
where $m_{\tilde{\chi}^+_1}<m_{\tilde{\chi}^+_2}$.

If $CP$ invariance is assumed, $U$ and $V$ are orthogonal matrices, whereas in $CP$-violating theories, the parameters $M_2$ and  $\mu$ can be complex. By a reparametrization of the fields, $M_2$ can be taken as real and positive \cite{Choi:1998ut}, so that the only non-trivial invariant phase is that of $\mu$, namely, $\mu=|\mu|e^{i\theta}$

The unitary matrices $U$ y $V$ must be chosen so that the elements of the diagonal matrix $M_{\tilde{\chi^+}}$ are real and non-negative. The corresponding eigenvalues are given by
\begin{equation}\label{charginomass}
m_{\tilde{\chi}_{1,2}^\pm}^2=\dfrac{1}{2}\Bigl(M_2^2+|\mu|^2+2\,m_W^2\mp \Delta\Bigr),
\end{equation}
where
\begin{equation}\label{deltamass}
\Delta=\sqrt{(M_2^2-|\mu|^2)^2+4\,m_W^4c^2_{2\beta}+4\,m_W^2(M_2^2+|\mu|^2+2 M_2|\mu|s_{2\beta} c_\theta)}.
\end{equation}

We can observe from Eqs. (\ref{charginomass}) and (\ref{deltamass}) that the chargino masses $m_{\tilde{\chi}_i^+}$ depend mainly on $\mu$, $M_2$ and $t_\beta$, which are constrained by the  LEP bound on the mass of the lightest chargino: $m_{\tilde{\chi}^\pm}>103.5\, \mbox{GeV}.$

\subsection{Sfermion sector}

In addition to $t_\beta$ and $\mu$, the sfermion sector can be described by three additional parameters for each sfermion kind: the left- and right-handed soft SUSY-breaking scalar masses $m_{\bar{f}_L}$ and $m_{\bar{f}_R}$ along with the trilinear coupling $A_f$. The sfermion mass matrix can be written as

\begin{equation}
\mathcal{M}_{\tilde{f}}^2=\left( \begin{array}{cc}
m_f^2+m_{LL}^2	&m_fX_f\\
m_F X_f	&m_f^2+m_{RR}^2
\end{array}\right),
\end{equation}
with
\begin{equation}
\begin{split}
m_{LL}^2&=m_{\tilde{f}_L}^2+(I_f^{3L}-Q_f s_W^2)m_Z^2 c_{2\beta},\\
m_{RR}^2&=m_{\tilde{f}_R}^2+Q_f s_W^2m_Z^2 c_{2\beta},\\
X_f&=A_f-\mu t_\beta^{-2I_f^{3L}}.
\end{split}
\end{equation}

In order to obtain the mass eigenstates  $\tilde{f}_1$ and $\tilde{f}_2$, the sfermion matrices are diagonalized by  a rotation by an angle $\theta_f$ via a $2\times 2$ unitary matrix
\begin{equation}
R^{\tilde{f}}=\left(\begin{array}{cc}
c_{\theta_f}	&c_{\theta_f}\\
-s_{\theta_f}    &c_{\theta_f}
\end{array}\right).
\end{equation}
The mixing angle and sfermion masses are then given by
\begin{align}
s_{2\theta_f}&=\frac{2 m_fX_f}{m_{\tilde{f}_1}^2-m_{\tilde{f}_2}^2},\\
c_{2\theta_f}&=\frac{m_{LL}^2-m_{RR}^2}{m_{\tilde{f}_1}^2-m_{\tilde{f}_2}^2},\\
m_{\tilde{f}_{1,2}}&= m_f^2+\frac{1}{2}\left[ m_{LL}^2+m_{RR}^2\mp \sqrt{(m_{LL}^2-m_{RR}^2)^2+4m_f^2 X_f^2}\right].
\end{align}
It is usually assumed that the masses of the first and second generation sfermions are zero \cite{Djouadi:2005gj}.

For the calculation of the  $\phi\to Z \gamma\gamma$ ($\phi=H,A$) decay  we find it convenient to work in the unitary gauge. The necessary Feynman rules are shown in Appendix \ref{FeynmanRules}. Below we present the analytical expression for the invariant amplitude and the corresponding decay width.

\section{  $\phi\to Z \gamma\gamma$ ($\phi=h,H,A$) decay width}

We turn to the most relevant aspects of our calculation. In the unitary gauge, the decay $\phi\to Z \gamma\gamma$ ($\phi=h,\,H,\,A$) is induced at the one-loop level by charged particles through the box and reducible Feynman diagrams shown in Figs. \ref{Abox} and \ref{reducibleset}. Apart from the loops with SM charged fermions and the $W$ gauge boson, in the MSSM there are contributions from  the charged scalar boson, charginos, and sfermions. Due to $CP$ invariance, the $W$ gauge boson, the charged scalar boson, and the sfermions only contribute through reducible diagrams.
%
%
%
%
%

We first define the kinematics conditions and introduce a set of invariant variables meant to simplify our analytical results. The nomenclature for the four-momenta of the external particles is as follows

\begin{equation}
\phi(p)\to \gamma_\mu(k_1)+\gamma_\nu(k_2) +Z_\alpha(k_3),
\end{equation}
with $p^2=m_\phi^2$, $k_1^2=k_2^2=0$ and  $k_3^2=m_Z^2$ due to the mass-shell condition. We now  introduce the following Lorentz invariant kinematic variables

\begin{equation}\label{scaledvariables}
\begin{split}
s_1&= (k_1+k_3)^2,\\
s_2&=(k_2+k_3)^2,\\
s&= (k_1+k_2)^2,
\end{split}
\end{equation}
which obey $s+s_1+s_2=m_\phi^2+m_Z^2$ by four-momentum conservation. All the scalar products between the external four-momenta can be expressed in terms of $s$, $s_1$ and $s_2$, together with the scaled variable $\mu_Z=m_Z^2/m_\phi^2$, so that the invariant amplitudes  of the $\phi\to Z\gamma\gamma$ $(\phi=h,H,A)$ decay can be written in terms of these variables. Furthermore, using the transversality condition for the external gauge bosons, i.e., $k_1\cdot \epsilon^\mu(k_1)=k_2\cdot \epsilon^\nu(k_2)=k_3\cdot \epsilon^\alpha(k_3)=0$, all the terms proportional to $k_1^\mu$, $k_2^\nu$ and $k_3^\alpha$ can be dropped from the invariant amplitude, which simplifies considerably the calculation.

After writing out the corresponding amplitude for each Feynman diagram, the one-loop integrals were reduced to scalar integrals via the Passarino-Veltman reduction scheme \citep{Passarino:1978jh}, which was carried out with the aid of the Mathematica package FeynCalc \cite{Mertig:1990an}. We first present the results for the $A\to Z\gamma\gamma$ decay.

\subsection{$A\to Z\gamma\gamma$ decay}
\subsubsection{Box diagram contribution}
\begin{figure}[ht!]
\centering
\includegraphics[width=0.65\textwidth]{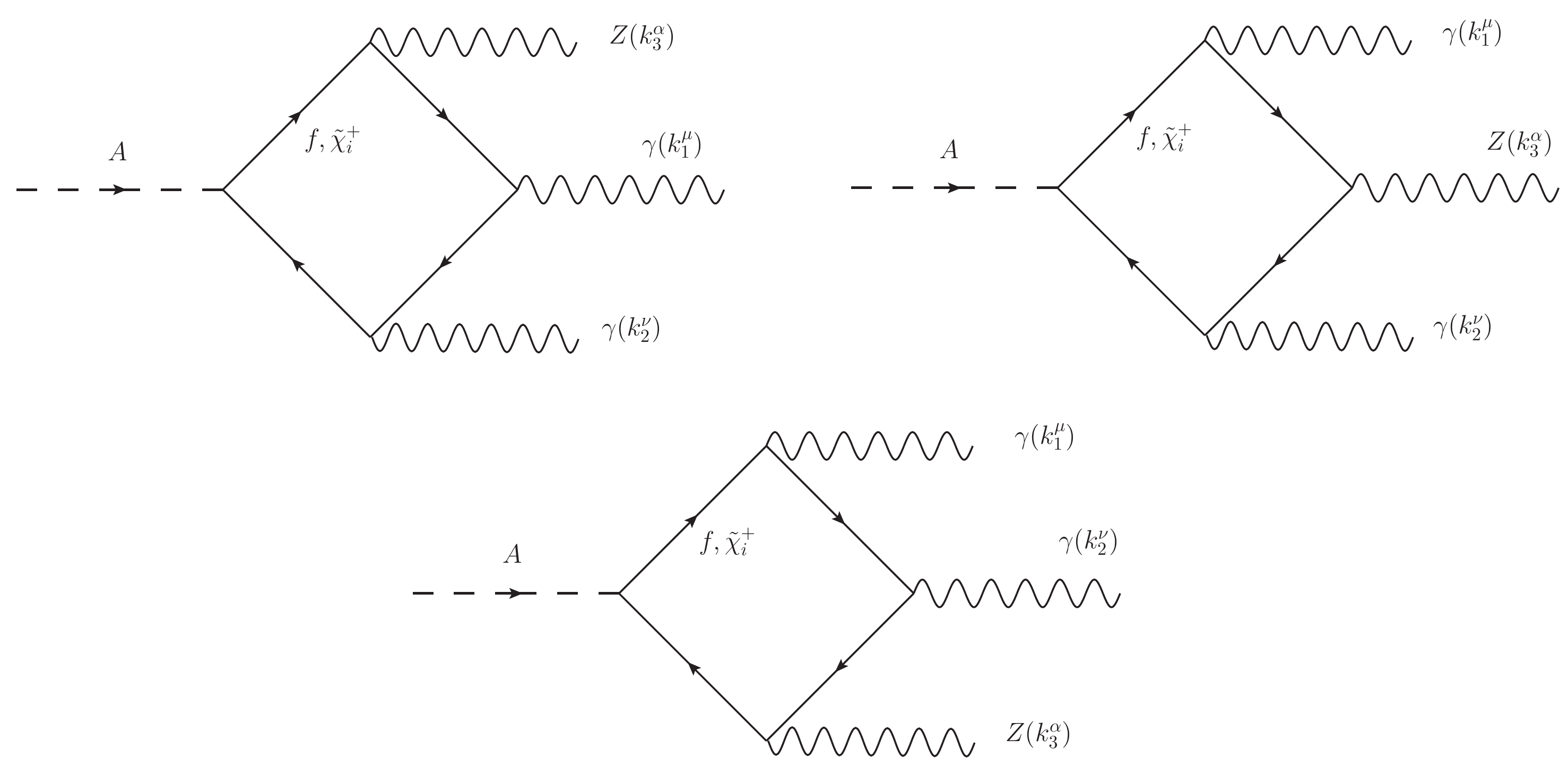}
\caption{Box diagrams that contribute to the  $A\to Z\gamma\gamma$ decay in the MSSM, with $f$ standing for any charged SM fermion  and  $\tilde{\chi}_{i}^+$ ($i=1,2$)  for two charginos. The crossed diagrams in which the photons are exchanged are not shown. The box diagrams for the process $\phi\to Z\gamma\gamma$ ($\phi=h,\,H$) are obtained by replacing $A$ by $\phi$.}
\label{Abox}
\end{figure}

This decay receive contributions of box diagrams and reducible diagrams. The  box diagrams are shown in Fig. \ref{Abox}:  the particles running into the loops are SM charged fermions and charginos. By $CP$ invariance, the pseudoscalar boson cannot couple to a pair of $W$ gauge bosons or charged scalar bosons. Also, it  cannot couple to a pair of identical sfermions by Bose symmetry. Therefore, the only new contribution from the MSSM to box diagrams  arises from charginos. Although there can be two charginos of distinct flavor into the box diagrams (there are non-diagonal couplings $Z\chi_1^+\chi_2^+$ and $A\chi_1^+\chi_2^+$ in the MSSM)  we only consider the contribution of a single chargino  as  non-diagonal chargino couplings are much more suppressed than the diagonal ones. In the case of the SM fermions,  the dominant contributions arise from the heaviest quarks: for small $t_\beta$, the top quark box dominates, whereas for large $t_\beta$ the bottom quark box  becomes relevant. This is due to the presence of the factors $1/t_\beta$ and $t_\beta$ in the corresponding Yukawa couplings.

Once the Passarino-Veltman reduction was applied to reduce the amplitude of each box diagram into  a combination of Passarino-Veltman scalar functions, we verified  that the total amplitude is ultraviolet finite and obeys $U(1)_{\mbox{em}}$ gauge invariance as well as Bose symmetry. Although each box diagram has ultraviolet divergences, they are cancelled out when summing over all diagrams. The full  amplitude can be arranged in the following manifestly gauge-invariant form

\begin{equation}
\mathcal{M}=\mathcal{M}^{\mu\nu\alpha}\epsilon^*_\mu(k_1)\epsilon_\nu^*(k_2)\epsilon_\alpha^*(k),
\end{equation}
with the Lorentz structures given as follows

\begin{equation}\label{amplitudeA}
\begin{split}
\mathcal{M}^{\alpha\mu\nu}&=\mathcal{F}_{1}^{\rm Box} \;k_1^\alpha\Bigl(k_1^\nu k_2^\mu-k_1\cdot k_2\; g^{\mu\nu}\Bigr)+\mathcal{F}_{2}^{\rm Box} \Bigl(k_3^\nu(k_1^\alpha k_2^\mu- k_1\cdot k_2\; g^{\alpha\mu})+k_2\cdot k_3(k_1^\nu g^{\alpha\mu}-k_1^\alpha g^{\mu\nu})\Bigr)\\
&+\mathcal{F}_{3}^{\rm Box} \;k_2^\alpha\Bigl(k_3^\mu(k_2\cdot k_3\;k_1^\nu-k_1\cdot k_2\; k_3^\nu)+k_1\cdot k_3(k_3^\nu k_2^\mu-k_2\cdot k_3\; g^{\mu\nu})\Bigr)+(k_1^\mu \leftrightarrow k_2^\nu).
\end{split}
\end{equation}
The form factors $\mathcal{F}_i^{\rm Box} $ depend on the variables $s$, $s_1$, $s_2$, and $\mu_Z$.  The explicit expressions  are somewhat cumbersome and are presented in Appendix \ref{FFactors}.

\subsubsection{Reducible diagram contribution}
\begin{figure}[ht!]
\centering
\includegraphics[width= 8cm]{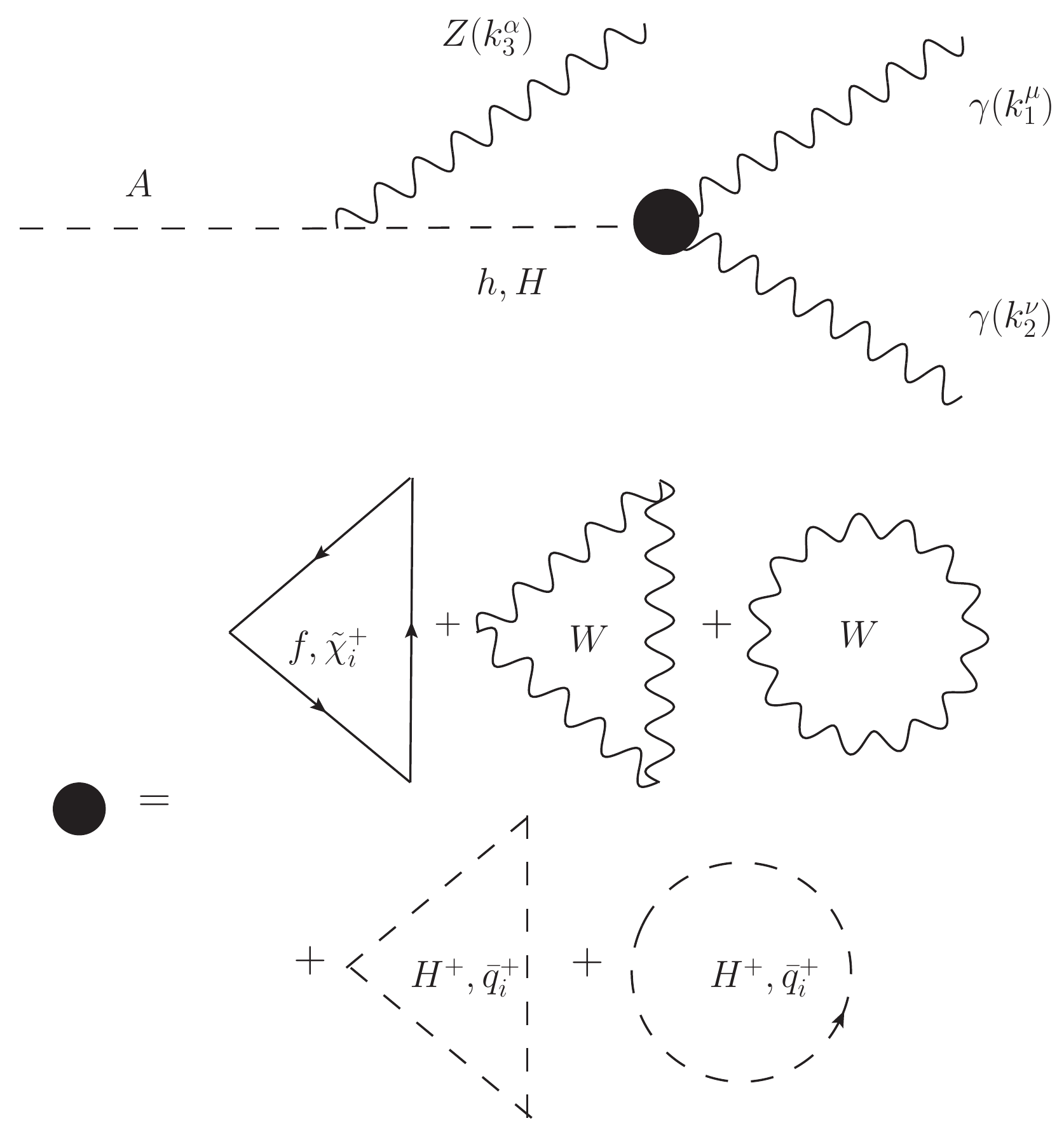}
\caption{Reducible Feynman diagrams for the $A\to Z\gamma\gamma$ process in the MSSM. There are additional triangle diagrams that can be obtained by exchanging the photons in the triangle diagrams. }
\label{reducibleset}
\end{figure}

The decay $A\to Z\gamma\gamma$ is also induced by the reducible Feynman diagrams shown in Fig. \ref{reducibleset}, which
are mediated by a  $CP$-even Higgs boson $\phi=h,H$ decaying into a photon pair via  triangle and bubble diagrams. The decay thus proceeds as follows $A\to Z\phi^*\to Z\gamma\gamma$. The  $AZZ$ coupling is absent at tree level and so are reducible diagrams mediated by a $Z$ gauge boson.

Reducible diagrams also yield an ultraviolet-finite and gauge invariant amplitude by themselves. They contribute to the general amplitude of Eq. \eqref{amplitudeA} through the form factor $\mathcal{F}_1$ via the loops with SM charged fermions, the $W$ gauge boson, charginos,  sfermions, and the charged scalar boson $H^\pm$. Note that two identical sfermions do couple to a $CP$-even Higgs boson and so they can contribute to the triangle diagrams. The triangle diagram contribution to the $\mathcal{F}_1$ form factor can be written as

\begin{equation}
\label{F1RD}
\mathcal{F}_1^{\rm RD}=\mathcal{F}_1^{\phi-\bar{\chi}_i^+}+\mathcal{F}_1^{\phi-\tilde{q}_i}+\mathcal{F}_1^{\phi-H^\pm}+\mathcal{F}_1^{\phi-f}+\mathcal{F}_1^{\phi-W},
\end{equation}
where $\phi=h,H$  and the  $\mathcal{F}_1^{\phi-\mathcal{X}}$ $(\mathcal{X}=f,W,H^\pm,\bar{\chi}_i^+,\tilde{q}_i)$ functions  are presented in Appendix \ref{FFactors} in terms of Passarino-Veltman scalar functions.

The total contribution to the $\mathcal{F}_i$ form factors is thus given by $\mathcal{F}_i=\mathcal{F}_i^{\rm Box}+\mathcal{F}_i^{\rm RD}$.

\subsection{$\phi\to Z\gamma\gamma$ ($\phi=h,H$) decay}

\subsubsection{Box diagram contribution}

These diagrams are similar to those shown in Fig. \ref{Abox}  and receive  contributions from SM charged fermions and charginos. Although there can be box diagrams with the $W$ gauge bosons, the charged scalar boson, and sfermions, these contributions exactly cancel out: it turns out that by  $CP$ invariance the  $\phi Z\gamma\gamma$ vertex function is composed by Levi-Civita tensors,  which can only arise from  the amplitudes of fermions loops involving a trace of a Dirac chain with a $\gamma^5$ matrix.

The most general Lorentz structure for the  $\phi\to Z\gamma\gamma$ ($\phi=h,H$) decay can be written in the following $U(1)_{\rm em}$ gauge-invariant manifest form

\begin{equation}
\begin{split}
\label{amplitudephi}
\mathcal{M}^{\alpha\mu\nu}(\phi\to Z\gamma\gamma)&=\mathcal{G}_1 \Big(k_1\cdot k_2\;\epsilon^{\alpha\mu\nu k_3}+g^{\mu\nu}\epsilon^{\alpha k_3k_1k_2}-k_2^\mu\epsilon^{\alpha\nu k_3k_1}+k_1^\nu\epsilon^{\alpha\mu k_3 k_2}\Big)\\
&+\frac{\mathcal{G}_2}{m_\phi^2} \;\epsilon^{\alpha\mu k_3 k_1}\Big(k_3\cdot k_2\; k_1^\nu-k_1\cdot k_2 \;k_3^\nu\Big)+ \mathcal{G}_3 \Big(k_1\cdot k_2\;\epsilon^{\alpha\mu\nu k_1}+k_1^\nu\epsilon^{\alpha\mu k_1 k_2}\Big)\\
&+\mathcal{G}_4 \Big(k_3\cdot k_2\; \epsilon^{\alpha\mu\nu k_1}+k_3^\nu\epsilon^{\alpha\mu k_1 k_2}\Big)+\Big(k_1^\nu\leftrightarrow k_2^\mu\Big),
\end{split}
\end{equation}
where we use the shorthand notation $\epsilon^{\alpha k  p q}=\epsilon^{\alpha\beta\lambda\rho}k_\beta p_\lambda q_\rho$, etc. To arrive to this expression, we used Schouten's identity.  The form factors $\mathcal{G}_i$, which depend on the kinematic variables $s$, $s_1$,  $s_2$, and $\mu_Z$,  are presented in Appendix \ref{FFactors} in terms of Passarino-Veltman scalar functions.

\subsubsection{Reducible diagram contribution}
\begin{figure}[ht!]
\centering
\includegraphics[width= 8 cm]{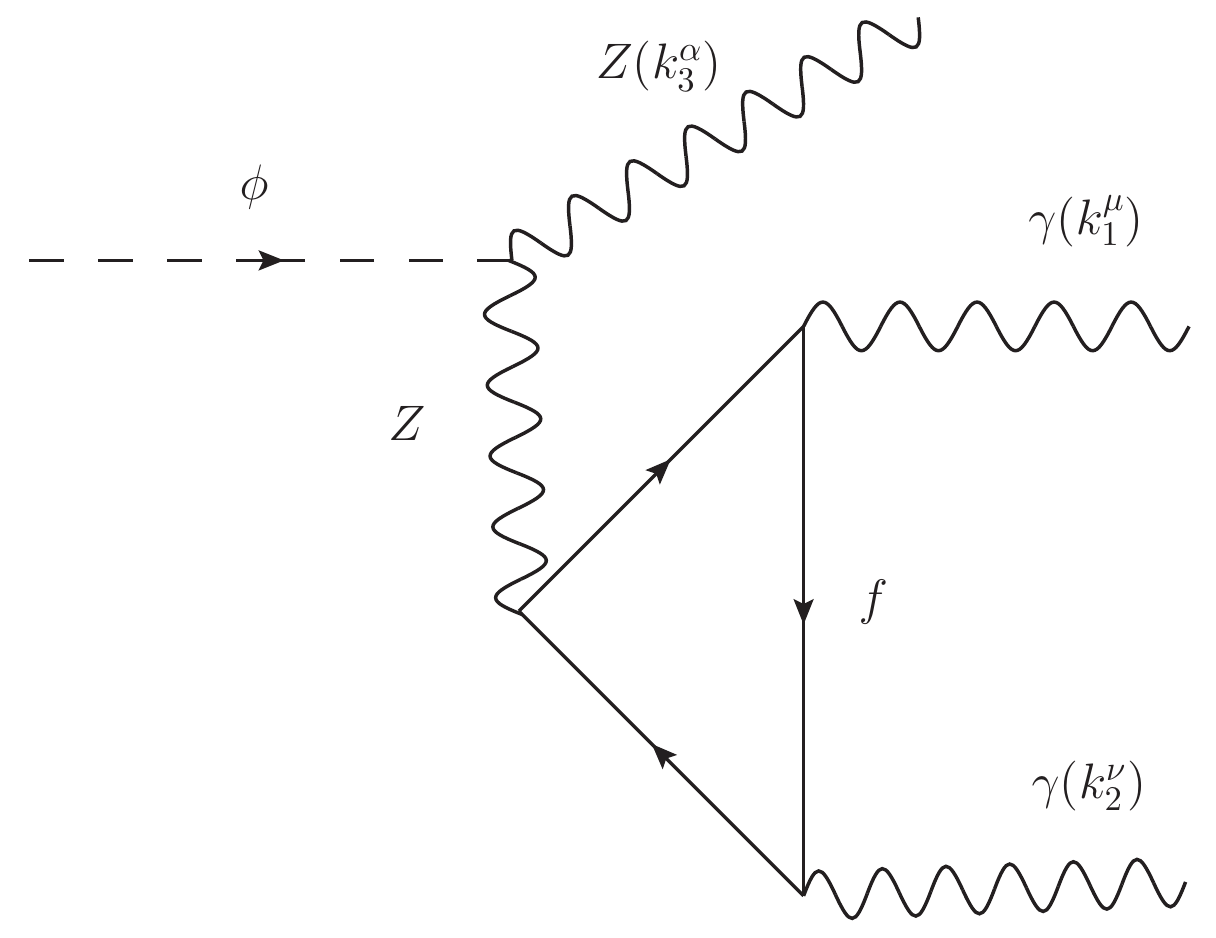}
\caption{Extra reducible Feynman diagram that contribute to the $\phi\to Z\gamma\gamma$ ($\phi=h,H$) decay in the MSSM. The diagram obtained by exchanging the photons is not shown.}
\label{reducibleZ}
\end{figure}

There are  two kinds of reducible Feynman diagrams. Those  of the first kind are similar to the ones shown in Fig. \ref{reducibleset}, but with the virtual  $CP$-even Higgs boson  replaced by the $CP$-odd Higgs boson $A$, which means that there are only loops with SM charged fermions and charginos. For the reasons  explained above, there are no triangle loops with virtual $W$ gauge bosons, charged scalar bosons, or sfermions. There is also another kind of reducible Feynman diagrams mediated by the $Z$ gauge boson as shown in Fig. \ref{reducibleZ} (the well known triangle anomaly $Z^*\gamma\gamma$), which  receive contributions from charged fermions only:  only fermion loops can give rise to the Levi-Civita tensor appearing in the corresponding vertex function via the trace of a Dirac chain with a $\gamma^5$ matrix.  $Z$-mediated reducible diagrams can also get chargino contributions, but they are  proportional to $g_A^{\tilde{\chi}_i}$, which exactly cancels in the hMSMM. Reducible diagrams only contribute to the  form factor $\mathcal{G}_3$, which reads

\begin{equation}
\label{GRD}
\mathcal{G}_3^{\rm RD}=\mathcal{G}_3^{A-f}+\mathcal{G}_3^{A-\tilde \chi^+}+\mathcal{G}_3^{Z-f},
\end{equation}
where $\mathcal{G}_3^{A-\mathcal{X}}$ ($\mathcal{X}=f,\tilde \chi^+$)  and $\mathcal{G}_3^{Z-f}$ stand for the contributions of the reducible Feynman diagrams mediated by the $CP$-odd  Higgs boson  and the $Z$ gauge boson, repectively. The corresponding expressions are presented in Appendix \ref{FFactors}.

\subsection{$\phi\to Z\gamma\gamma$ ($\phi=h,H,A$) decay width}
The square of the  invariant amplitudes of Eqs. \eqref{amplitudeA} and \eqref{amplitudephi}, averaged over polarizations of the photons and the $Z$ gauge boson, can be  obtained after some lengthy algebra. The results are shown in Appendix \ref{squaredamplitude} and can be plugged into the following formula to obtain the decay width

\begin{equation}\label{decaywidth}
\Gamma(\phi\rightarrow Z\gamma\gamma) =\dfrac{m_\phi}{256\; \pi^3}\int_{x_{1i}}^{x_{1f}} \int_{x_{2i}}^{x_{2f}} |\bar{\mathcal{M}}(\phi\to Z\gamma\gamma)|^2 dx_2\;dx_1,
\end{equation}
for $\phi=h,H,A$. In the rest frame of the decaying Higgs boson, the scaled variables $s,s_1$ and $s_2$, defined in Eq. \eqref{scaledvariables}, are related to the energy of the $Z$ boson (one of the photons) $E_Z$  ($E_\gamma$) according to

\begin{align}
\label{svariable}
s&=m_\phi^2 (\mu_Z-x_1+1),\\
\label{s1variable}
s_1&=m_\phi^2(1-x_2),\\
\label{s2variable}
s_2&=m_\phi^2(x_1+x_2-1),
\end{align}
where $x_1=2E_Z/m_\phi$ and $x_2=2E_\gamma/m_\phi$. The kinematic limits of Eq. \eqref{decaywidth} read

\begin{align}
2\sqrt{\mu_Z}&\le x_1\le 1+\mu_Z,\\
x_{2}&\gtreqless\dfrac{1}{2}\left (2-x_1\mp \sqrt{x_1^{ 2}-4\mu_Z}\right ).
\end{align}

Below we analyze the behavior of the $\phi\to Z\gamma\gamma$ ($\phi=h,H,A$)  decay width in the parameter space of the MSSM still allowed by current constraints.

\section{Numerical Analysis and Results}
 The MSSM parameters that are relevant to our study are mostly associated to the electroweakino  and Higgs sectors, namely,  $M_2$, $\mu$, $m_A$,  $t_\beta$, as well as the sfermion and chargino masses. We first present an overview of the allowed values of these parameters,  according to experimental and theoretical constraints, focusing on the so called hMSSM scenario. Afterwards we evaluate the $\phi\to Z\gamma\gamma$ ($\phi=h,H,A$) decay in the allowed region of the parameter space.

\subsection{Allowed parameter space of the hMSSM}

Until now, most of the observed effects at  high energy experiments have been found compatible with the SM: the experimental data  agree with the theoretical predictions to a high accuracy. The status of the SM was cemented by the Higgs boson discovery at the LHC  \cite{Aad:2015zhl}. Subsequent measurements   \cite{Khachatryan:2016vau} leave small room for NP effects and so can have strong implications on the parameter space of BSMTs such as the MSSM. In particular, the parameters associated with the scalar sector of the MSSM have become tightly constrained by the measurement of the couplings of  the 125 GeV Higgs boson,  whose properties   must be reproduced by the lightest $CP$-even Higgs boson $h$ of the MSSM. This favors regions very close  to the alignment limit, $\sin(\beta-\alpha)=1$, in which the $h$ couplings to SM particles are SM-like. In such a  limit, the mass eigenbasis of the $CP$-even sector coincides with the one in which the gauge bosons receive their masses from only one of the two Higgs doublets. Also, the parameters $\alpha$ and $\beta$ are  related by  $\tan\alpha=-1/\tan\beta$,  and the couplings of the heavy $CP$-even Higgs boson to the weak gauge bosons vanish,  although it can still couple to the fermions. Furthermore, the non-observation of any additional Higgs bosons at the LHC implies that their masses are likely to lie above the elecroweak scale i.e. $m\phi \gg m_Z$.  For our calculation we consider a region close to the decoupling limit of the MSSM \cite{Djouadi:2005gj} in which $\sin(\beta-\alpha)=1$ and consider that the new Higgs bosons are heavy.

\subsubsection{{\rm SUSY} particle masses}
The discovery of the Higgs boson at the LHC also has had severe implications on the sfermion sector as the stop mass is required to be of the order of 1--10 TeV to accommodate a 125 GeV scalar boson. However, the  tension between the  experimental and theoretical values of the muon anomalous magnetic moment can be alleviated  by the presence of SUSY particles with masses at the GeV scale, which are also required by  the little hierarchy problem. This suggests that SUSY particles such as electroweakinos may have masses  at the GeV scale. The search for such particles is one of the main goals of the LHC after the Higgs discovery. Along this line, the non-observation of squarks and gluinos at the early stage of the LHC,  hints that the masses of such particles lie in the TeV scale. Also, direct electroweakino searches have imposed strong bounds on their masses  \cite{Aad:2015eda,Aad:2014yka,Aad:2015jqa} via the chargino-neutralino pair production channel. The most up-to-date analysis constrains  the  lightest neutralino mass to the  $100-150$ GeV interval, whereas the lower bound on the mass of a degenerate wino-like $\chi_2^0$ and $\chi_1^\pm$ is about $300$ GeV. In view of these constraints, a heavy  Higgs boson with a mass below $500-600$ GeV is not kinematically allowed to decay into an  electroweakino final state.

\subsubsection{$CP$-odd Higgs scalar boson mass $m_A$ and  ratio of VEVs $t_\beta$}

Direct searches of heavy scalar resonances at the LHC are useful to contrain  $m_A$ and $t_\beta$. Recently, the  combined LHC data at $\sqrt{s}=7$ and 8 TeV allowed the ATLAS and CMS Collaborations searching for neutral scalar bosons and constraining the parameter space of the MSSM  via the following decay channels $H\to \gamma\gamma$ \cite{Aaboud:2017yyg}, $H\to WW$ \cite{Aad:2015agg}, $H\to ZZ$ \cite{Aad:2015kna} $H\to hh\to b\bar{b}b\bar{b}$, $H\to hh\to b\bar{b}\gamma\gamma$ \cite{Aaboud:2018ftw}, $ A\to Zh$ \cite{Aad:2015wra} and $H/A\to \tau^+\tau^-$ \cite{Khachatryan:2014wca}, with the latter  channel imposing the most stringent bounds. The most up-to-date constraints arise from the analysis of the LHC data for Higgs boson production in association with  bottom quarks $\bar{b}bH/A$ and via gluon fusion at $\sqrt{s}=13$ TeV  \cite{Aaboud:2017sjh,Sirunyan:2018zut}.  Such data were used by the ATLAS Collaboration to perform a search for neutral Higgs bosons  in the $200$--$1200$ GeV mass range decaying into the $\tau^+\tau^-$ channel \cite{Aaboud:2017sjh}. It was found that for $m_A=200$ GeV (1500 GeV), the region with $t_\beta>1$ ($t_\beta>42$) is excluded at 95\% C.L. A similar study by the CMS Collaboration \cite{Sirunyan:2018zut}, yields slightly less stringent constraints. Other decay channels such as $H\to ZZ$ and $A\to hZ$, are useful to exclude the region with $t_\beta\lesssim 4$ and $m_A\lesssim 350$ GeV \cite{Aad:2015pla}, but the region with $m_A\gtrsim 400$ GeV and $t_\beta\lesssim 10$ is still allowed.

As for indirect constraints on the MSSM parameters, the most stringent bounds can be obtained from the study of  the rare $B$-meson decays $B_s\to X_s\gamma$ and $B_s\to \mu^+\mu^-$, whose branching ratios  are constrained to the following 95\% C.L. intervals \cite{Amhis:2014hma}
\begin{align}
2.82\times 10^{-4}<{\rm BR}(B_s&\to X_s\gamma)<4.04\times 10^{-4},\\
1.57\times 10^{-9}<{\rm BR}(B_s&\to \mu^+\mu^-)<4.63\times 10^{-9}.
\end{align}
which allows one to  find the allowed region in the $m_A-t_\beta $ plane \cite{Bhattacherjee:2015sga}. It turns out that the region with $m_A<350$ GeV and $t_\beta\geq 25$ is excluded by  the $B_s\to \mu^+\mu^- $ decay, which is sensitive to the region with large $t_\beta$ and relatively small $m_A$. On the other hand, the region where $m_A<350 $ GeV and $t_\beta\leq 8$ is not favored by the $B_s\to X_s\gamma$ decay, whereas the region with large $m_A$ ($m_A\gtrsim 400$ GeV) and small or moderate values of $t_\beta$ is still unconstrained by the experimental data.

In our analysis we consider values of $t_\beta$ and $m_A$ in the following intervals

\begin{equation}
1< t_\beta < 15\;\;\; {\rm and}\;\;\; 400\; \mbox{GeV}<m_A<750\; \mbox{GeV}.
\end{equation}
\subsubsection{{\rm SUSY} parameters}
The input parameters  $M_2$ and $\mu$ are relatively close and hence the charginos  are mixed states of bino  and Higgsino states.  We fix these values according to the current bounds. In particular, we take $M_2=500$ GeV and $\mu=400$ GeV in our analysis. As far as the $CP$-violating phase angle $\Phi_\mu$ we take $\cos(\Phi_\mu)=1$. With these input parameters, the  mass for the lightest  chargino is  $m_{\tilde{\chi}_1^+}=377.9$ for $t_\beta=5$. Finally, the $U(1)_Y$  gaugino mass parameter, $M_1$, is fixed via the GUT relation

\begin{equation}
M_1=\frac{5 s_W^2}{3 c_W^2}M_2.
\end{equation}

\subsection{Behavior of the decay $\phi\to Z\gamma\gamma$ ($\phi=h,H,Z$)  in the hMSSM}
We first present a general discussion of  the most interesting features of the decay $\phi \to Z\gamma\gamma$ ($\phi=h,H,Z$).
As already mentioned, the decay $A\to Z\gamma\gamma$ receives contributions from loops with SM charged fermions, the  $W$ gauge  boson, the charged scalar boson,  charginos, and squarks. On the other hand, only charged fermion and charginos contribute to the $\phi\to Z\gamma\gamma$  ($\phi=h,H$) decay. It turns out that the charged scalar contribution to the  $A\to Z\gamma\gamma$ decay is considerably smaller than the fermion contribution and so is the  $W$ gauge boson contribution, which is negligible in regions close to  the alignment limit as it is proportional to $c_{\beta-\alpha}$. Also,   squark contributions to  both $A\to Z\gamma\gamma$ and $\phi\to Z\gamma\gamma$  ($\phi=h,H$) decays are suppressed, which means that the only relevant contributions to both decays are those of SM charged fermions and charginos. Since fermion contributions are   proportional to the charged particle mass, the main contributions of SM fermions  arise from the top and bottom quarks. For low values of $t_\beta$ the most important contribution comes from the top quark, whereas for large $t_\beta$ the bottom quark contribution becomes relevant. As for the chargino contribution,  because of the presence of non-diagonal couplings of both the scalar boson $\phi$ and the $Z$ gauge boson to charginos, the box diagrams for the decay $\phi\to Z\gamma\gamma$ ($\phi=h,H$) can include two distinct charginos into the loops.  However, non-diagonal chargino couplings are very suppressed and we can safely neglect this type of contributions, which in fact are absent in the reducible diagrams. Another point worth mentioning is that the calculation of  chargino box diagrams seems to be too cumbersome as the coupling $A\tilde{\chi_i}^+\tilde{\chi_j}^+$ ($h/H\tilde{\chi_i}^+\tilde{\chi_j}^+$)   is both scalar and pseudoscalar, as shown in Table  \ref{charginocouplings}, however,  the scalar (pseudoscalar) part vanishes in the SUSY limit. Therefore  the corresponding Lorentz structures for the chargino contributions are identical to those  induced by SM charged fermions.

 In THDMs the decay $A\to Z\gamma\gamma$  can have a considerable enhancement \cite{Sanchez-Velez:2018xdj} in the region of the parameter space where $m_A>m_Z+m_H$,  when the decaying $CP$-odd scalar boson is heavy enough to decay as $A\to ZH$, with the $CP$-even scalar boson decaying subsequently into a photon pair  $H\to\gamma\gamma$. A similar argument goes for the enhancement of the decay $H\to Z\gamma\gamma$ in the scenario with $m_H>m_Z+m_A$. However, neither scenario is possible in the MSSM as the masses $m_H$ and $m_A$ are not independent (they are   almost degenerate indeed) so the decays $A\to ZH$ and $H\to AZ$ are not kinematically allowed. Although the decay $A\to Zh$ is not kinematically forbidden for very heavy $m_A$,  it is very suppressed as its invariant amplitude is proportional to $c_{\beta-\alpha}$, which vanishes in the alignment limit. We thus do not expect any   enhancement of  the decay $\phi\to Z\gamma\gamma$ ($\phi=H,A$), let alone the decay $h\to Z\gamma\gamma$. The latter has a negligibly small branching fraction with no considerable devation from the SM prediction, of the order of $10^{-11}$, and so we refrain from analyzing it more detailed. We thus focus our study on the decays   $A\to Z\gamma\gamma$ and $H\to Z\gamma\gamma$.

Below we present the numerical analysis of  the $\phi\to Z\gamma\gamma$ ($\phi=H,A$) decay. The Passarino-Veltman scalar function were evaluated via the LoopTools routines \cite{Hahn:1998yk,vanOldenborgh:1989wn}. For the sake of completeness we also calculate  the main tree-level and one-loop level induced two body-decays  into SM particles, such as $\phi \to \bar{b}b$, $\bar{t}t$, $\gamma\gamma$, $gg$, and $Z\gamma$. Notice that decays such as $A\to Zh$, $H\to ZZ$ and $H\to WW$ have negligible rates in regions of the parameter space close to the alignment limit. For comparison purposes  we also calculate the three-body decays $\phi\to Zt\bar{t}$ and $Zb\bar{b}$. As far as the decays into SUSY particles are concerned, the only kinematically allowed channel is the one into a neutralino pair $\phi \to\chi_1^0\chi_1^0$, which can give a relevant  contribution to the total width and was already  studied in \cite{Barman:2016kgt}. Other decays of the heavy neutral scalar bosons into SUSY particles such as   squarks or  charginos  are not kinematically allowed in the region of the parameter space we are considering.

\subsubsection{$A\to Z\gamma\gamma $ branching ratio}

\begin{figure}[htb!]
\centering
\includegraphics[width=.95\textwidth]{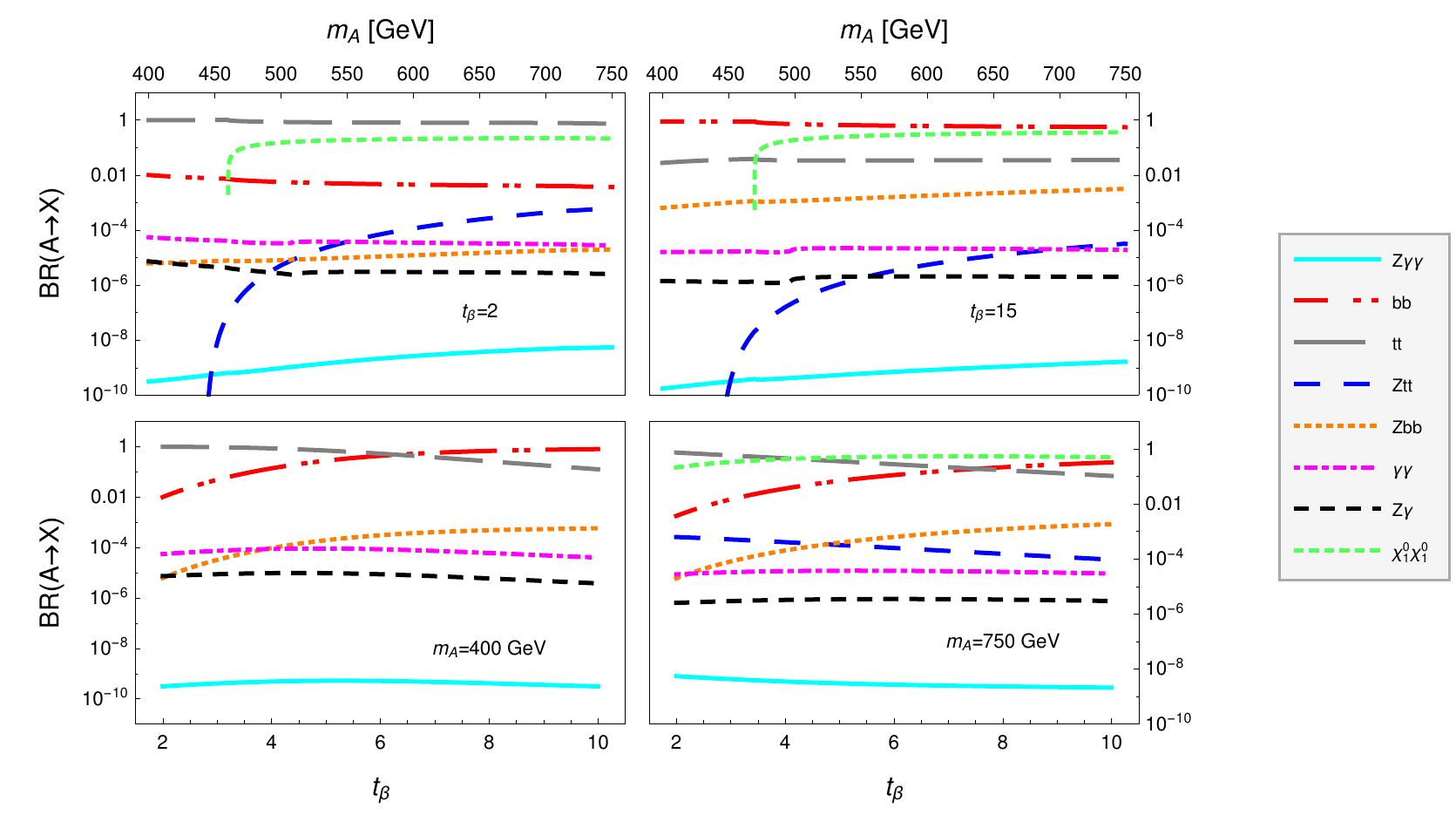} 
\caption{Branchig ratio for the $A\to Z\gamma\gamma$ decay in the hMSSM as a function of $m_A$ for two values of $t_\beta$ (top plots) and as a function of $t_\beta$ for two values of $m_A$ (bottom plots). We also include the dominant decay channels of the $CP$-odd scalar boson. The neutralino mass  $m_{\chi^0_1}$ is about 230 GeV.
\label{AplothMSSM}}
\end{figure}

In the region close to the decoupling limit, the  $AhZ$ coupling is highly suppressed, thus the reducible diagram contribution to the $A\to Z\gamma\gamma $ decay arises only from $H$ exchange, which in fact dominates  over the box diagrams contribution by almost two orders of magnitude. The behavior of the branching ratio ${\rm BR}(A\to Z\gamma\gamma)$ as a function of $m_A$ in the mass range 350-750 GeV is shown in the top plots of Fig. \ref{AplothMSSM} for  two values of $t_\beta$, whereas the bottom plots show its behavior as a function of $t_\beta$ for two values of  $m_A$. We can observe that for $t_\beta=2$ (top-left plot), ${\rm BR}(A\to Z\gamma\gamma)$ is of the order of $10^{-8}$ for $m_A$ below 350 GeV, but drops by more than one order of magnitude when the $\bar{t}t$ channel becomes open, though it increases again up to around $10^{-8}$ as  $m_A$ increases up to 750 GeV.  When  $t_\beta$ increases up to $15$ (top-right plot), there is no significant change in the ${\rm BR}(A\to Z\gamma\gamma)$ behavior, except around $m_A=350$ GeV when it is considerably smaller than in the $t_\beta=2$ scenario. In the bottom plots of Fig. \ref{AplothMSSM} it is evident that there is very little dependence of ${\rm BR}(A\to Z\gamma\gamma)$ on $t_\beta$ for $m_A>350$ GeV. We conclude that the rate for the three-body decay  $A\to Z\gamma\gamma$ is well below   those for the decays $A\to Z\gamma$ and $A\to\gamma\gamma$: $BR(A\to Z\gamma\gamma)/BR(A\to\gamma\gamma)\sim 10^{-4}$ and $BR(A\to Z\gamma\gamma)/BR(A\to Z\gamma)\sim 10^{-3}$  when $m_A=750$ GeV and   $t_\beta $ is  small.  These values decrease by at least one order of magnitude for $m_A$ around 400 GeV.

\subsubsection{$H\to Z\gamma\gamma $ branching ratio}
\begin{figure}
\centering
\includegraphics[width=.95\textwidth]{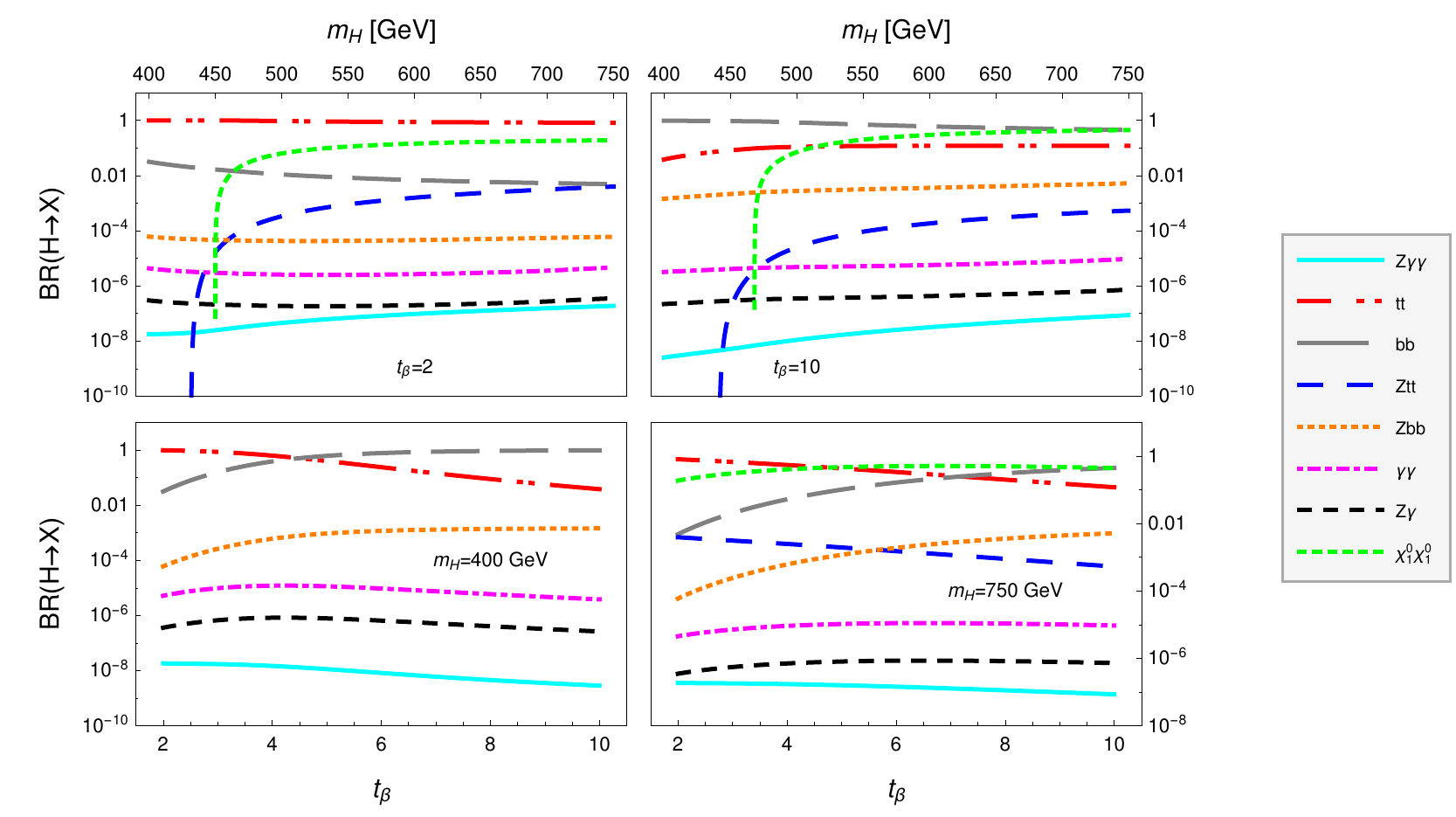} 
\caption{Branchig ratio for $H\to Z\gamma\gamma$ decay in the MSSM as a function of $m_A$ for two values of $t_\beta$ and as a function of $t_\beta$ for two values of $m_A$. The neutralino mass  $m_{\chi^0_1}$ is about 230 GeV.}
\label{HplothMSSM}
\end{figure}

We now perform an analysis of the behavior of the $H\to Z\gamma\gamma$ branching ratio. As in the case of the $A\to Z\gamma\gamma$ decay, box diagrams  give a smaller contribution to the $H\to Z\gamma\gamma$ decay than the ones arising from the reducible diagrams, which in this case can be mediated by a virtual $CP$-odd scalar boson and the $Z$ gauge boson. However, $Z$-mediated reducible diagrams give a negligible contribution since its amplitude is proportional to $c_{\beta-\alpha}$. Thus the main contribution to this decay arises from the reducible diagram with $H$ exchange  via the triangle loops with SM charged fermions and charginos. In the top plots of Fig. \ref{HplothMSSM} we show the branching ratio for the decay $H\to Z\gamma\gamma$ as a function of $m_H$ for  two values of $t_\beta$, whereas in the bottom plots  we shown ${\rm BR}(H\to Z\gamma\gamma)$ as a function of $m_H$ for two values of $m_A$. For $t_\beta=2$, ${\rm BR}(H\to Z\gamma\gamma)$ is of the order of $10^{-8}$ ($10^{-7}$) for  $m_A=350$ GeV  (750 GeV). These values are slightly larger than  those reached by ${\rm BR}(A\to Z\gamma\gamma)$. In fact for $t_\beta=2$ and large $m_H$, the rate for the three-body decay  $H\to Z\gamma\gamma$ is almost as large as the two-body decay $H\to Z\gamma$. However, for  $t_\beta=15$,  ${\rm BR}(H\to Z\gamma\gamma)$ decreases by about one order of magnitude: for instance, it is about $10^{-8}$ for $m_H=750$ but decreases by about one order of magnitude for smaller values of $m_H$. Therefore ${\rm BR}(H\to Z\gamma\gamma)$ is more sensitive to $t_\beta$ than ${\rm BR}(A\to Z\gamma\gamma)$. This fact becomes evident in the bottom plots of Fig. \ref{HplothMSSM}, where we observe that ${\rm BR}(H\to Z\gamma\gamma)$ decreases by almost one order of magnitude as $t_\beta$ increases from 2 to 15.

As far as the dominant decays are concerned, the  $\bar{t}t$ ($\bar{b}b$) channel is the dominant one for small (large) $t_\beta$. The only decay into SUSY particles,  the neutralino channel, can also be relevant once it becomes open. In fact for $m_A=400$ GeV, the rate  of the $\tilde\chi^0_1 \tilde \chi^0_1$ channel surpasses those of both the $\bar{t}t$ and $\bar{b}b$ channels in a narrow region around $t_\beta=6$, which becomes wider, about $5\lesssim t_\beta\lesssim 11$, for $m_A=750$ GeV.

\section{Conclusions}
In this work we have  presented an explicit calculation of the decays of the $CP$-odd and $CP$-even scalar bosons $A\to Z\gamma\gamma$ and $H\to Z\gamma\gamma$ in the framework of the MSSM, focusing on the so-called hMSSM scenario. These decays have been computed previously in the context of THDMs, where the main contributions arise from loops with the top and bottom quarks.
In the MSSM there are new contributions from squarks and  charginos. The latter is comparable to those of the top and bottom quarks for certain  values of the parameters of the electroweakino sector still consistent with the experimental bounds. As for the contribution from  squarks, it is negligibly small since the squark masses lie in the TeV scale. Although both decays  receive contributions from  box and reducible diagrams, the dominant contribution arises from the latter. We have presented numerical results for the corresponding branching ratios in the context of the decoupling limit  and considered  values for the free parameters $m_A$ and $t_\beta$ still consistent with constraints from experimental data. For the $A\to Z\gamma\gamma$ decay, we found that the branching ratio is suppressed by three orders of magnitude compared with the diphoton channel.  In general ${\rm BR}(A\to Z\gamma\gamma)$ increases as $m_A$ increases, though  it is not very sensitive to $t_\beta$. For $m_A=750$ GeV, ${\rm BR}(A\to Z\gamma\gamma)$ is of the order of $10^{-8}$ although it  can drop up to $10^{-10}$ for small values of $m_A$. As far as the $H\to Z\gamma\gamma$ decay is concerned, its branching ratio exhibits a similar behavior to that of the $A\to Z\gamma\gamma$ decay, although it decreases as $t_\beta$ increases, which stems from the fact that the coupling of the $CP$-odd Higgs boson to charginos is more suppressed  than that of the $CP$-even Higgs boson. It was found that ${\rm BR}(H\to Z\gamma\gamma)$ is of the order of $10^{-7}$ for $m_A=700$  GeV and $t_\beta=2$. For these values of the free parameters, ${\rm BR}(H\to Z\gamma\gamma)$ is close to   ${\rm BR}(H\to Z\gamma)$. As far as the MSSM contribution to the $h\to Z\gamma\gamma$ decay, it is negligible and the corresponding branching ratio is of the order of $10^{-11}$. Although the studied decays have very suppressed branching ratios,  the  $AZ\gamma\gamma$ and $HZ\gamma\gamma$ couplings can be of interest in the  study of heavy Higgs boson production at a linear collider working in the $\gamma\gamma$ mode.

\acknowledgments{We acknowledge support from Consejo Nacional de Ciencia y Tecnolog\'ia and Sistema Nacional de Investigadores. Partial support from Vicerrector\'ia de Investigaci\'on y Estudios de Posgrado de la Ben\'emerita Universidad Aut\'onoma de Puebla is also acknowledged. }
\appendix

\section{ Feynman rules for the decay $\phi\to Z\gamma\gamma$ ($\phi=h,H,A$) in the MSSM}
\label{FeynmanRules}
We now present all the Feynman rules necessary for the calculation of the MSSM contributions to the decay $\phi\to Z\gamma\gamma$ ($\phi=h,H,A$) in the unitary gauge.

\begin{figure}[ht!]
 \includegraphics[width= 10cm]{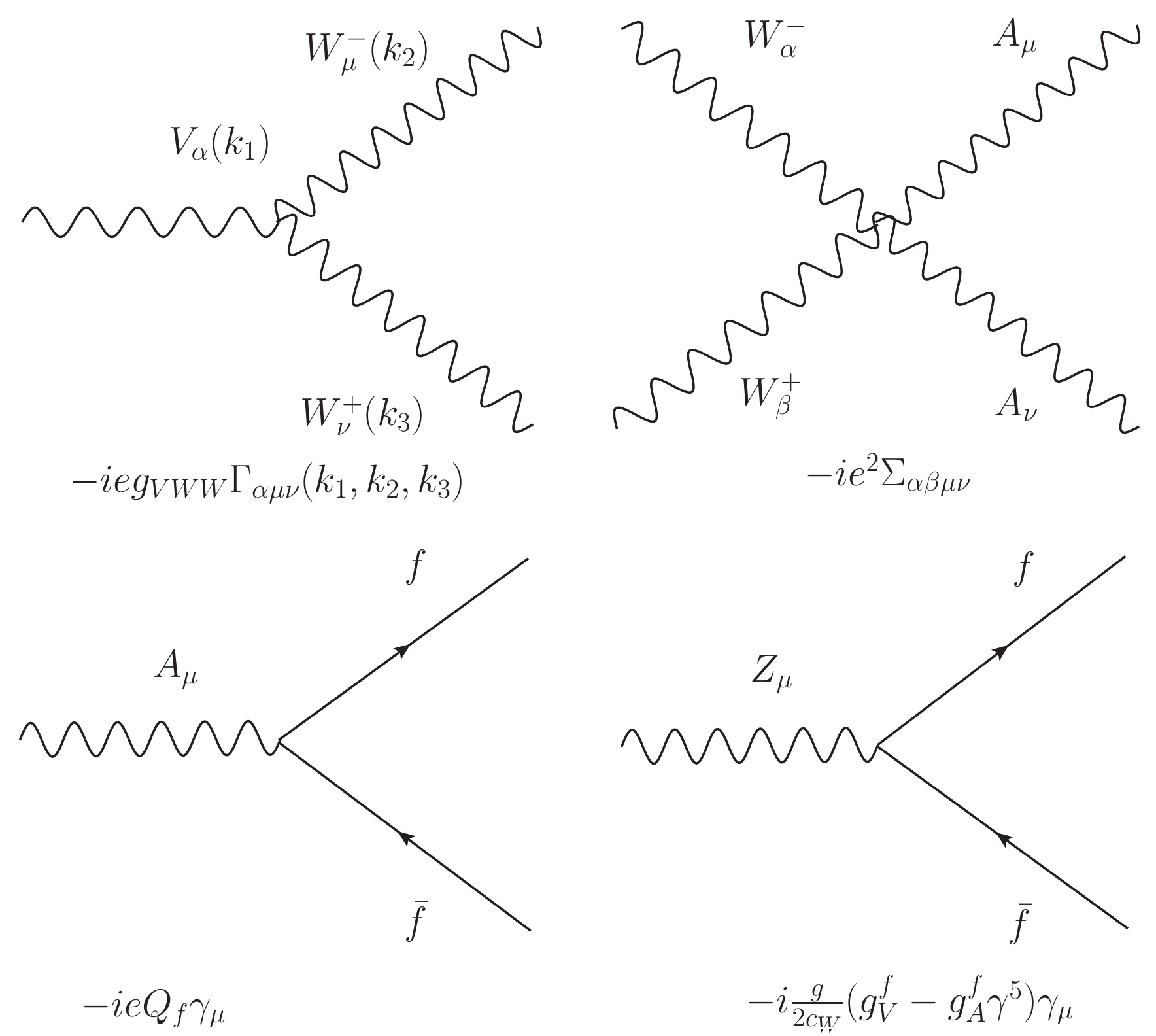}
 \caption{ Feynman rules involving only SM particles. All the 4-momenta are incoming and the Lorentz structures are given by $\Gamma^{\alpha\mu\nu}(k_1,k_2,k_3)=(k_1-k_2)^\nu g^{\alpha\mu}+(k_2-k_3)^\alpha g^{\mu\nu}+(k_3-k_1)^\mu g^{\alpha\nu}$ and $\Sigma^{\alpha\beta\mu\nu}=2g^{\alpha\beta}g^{\mu\nu}-g^{\alpha\mu}g^{\beta\nu}-g^{\alpha\nu}g^{\beta\mu}$. The coupling constants read $g_{VWW}=1$ ($-\frac{c_W}{s_W}$) for $V=\gamma$ $(Z)$ as well as  $g_A^f=\frac{1}{2}T^3_f$ and  $g_V^f=T_3^f-2Q_f s_W^2$, with $Q_f$ the fermion charge and $T^3_f=1/2$ ($-1/2$) for up quarks (down quarks and charged leptons).  \label{SMFeynmanRules}}
\end{figure}

 We first present in Fig. \ref{SMFeynmanRules} the Feynman rules for the following vertices involving only SM particles: $VW^-W^+$, $\gamma\gamma W^-W^+$, and $V\bar{f}f$ ($V=\gamma,Z$), which are the same as in the SM.

 We also need the interaction of the photon and the $Z$ gauge boson with a pair of charginos,  which are given as follows:
 \begin{align}
 A_\mu \bar\chi_i^+\chi_j^- :&\quad-ie\delta^{ij}\gamma_\mu,\\
 Z_\mu \bar\chi_i^+\chi_j^- :&\quad \frac{-ie}{4 c_W s_W}\gamma_\mu(g_V^{\tilde{\chi}_{ij}}-g_A^{\tilde{\chi}_{ij}}\gamma^5).
 \end{align}
where the indices $i,j$ stand for the chargino generation, and the coupling constants are given by
\begin{align}
g_V^{\tilde{\chi}_{ij}}&=V_{i1}V_{j1}^*+U_{i1}^*U_{j1}+2\delta_{ij}(c_W^2-s_W^2),\\ g_A^{\tilde{\chi}_{ij}}&=V_{i1}^*V_{j1}^*-U_{i1}^*U_{j1}.
\end{align}

\begin{figure}[ht!]
\centering
\includegraphics[width= 13 cm]{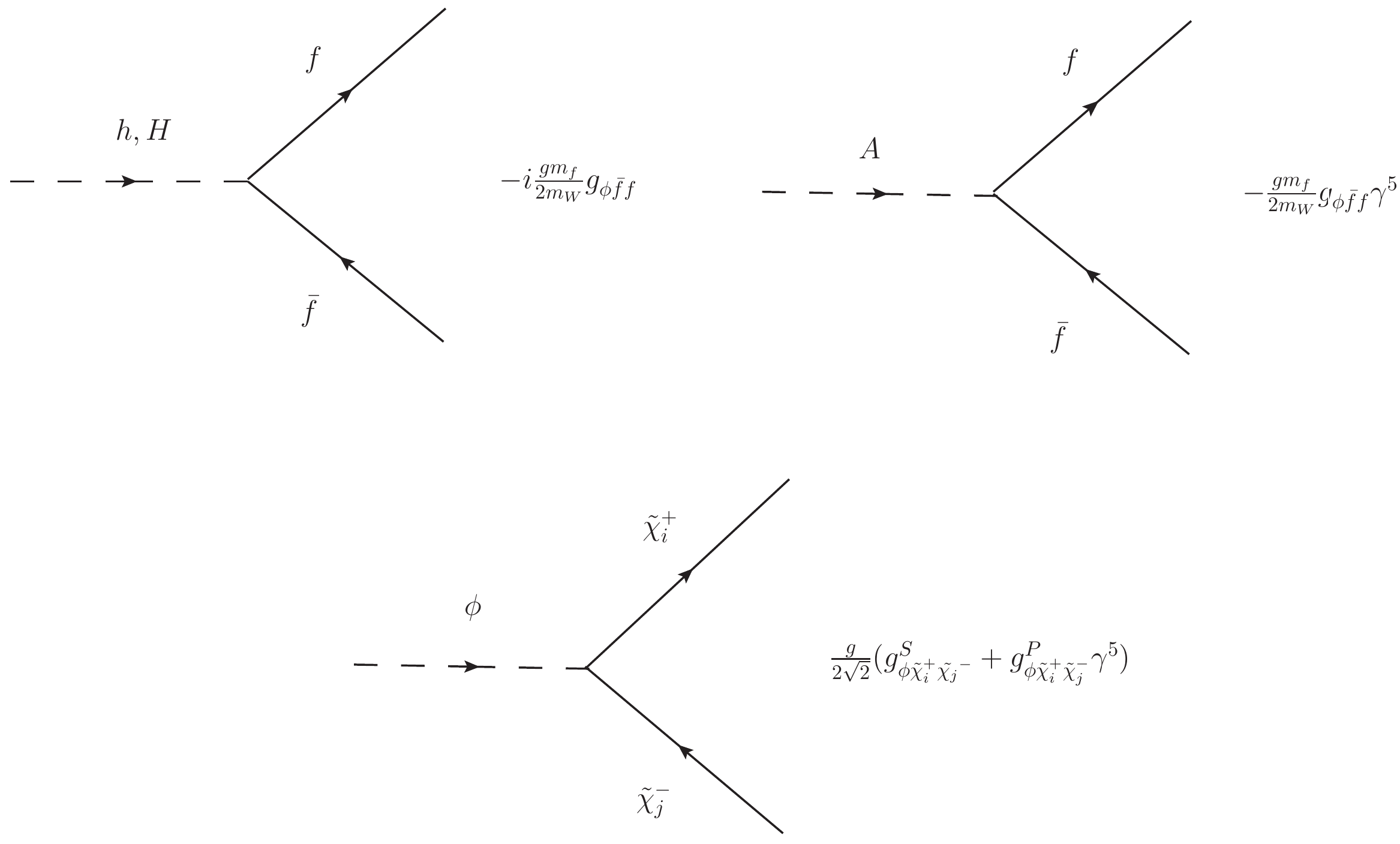}
\caption{Feynman rules for the couplings of the scalar bosons to fermions and charginos in the MSSM. In the lower diagram $\phi=h,H,A$,  and the corresponding coupling constants are shown in Tables \ref{CouplingConstants2} and \ref{charginocouplings}. }
\label{FRscalartofermions}
\end{figure}

\begin{figure}[ht!]
\centering
\includegraphics[width= 13 cm]{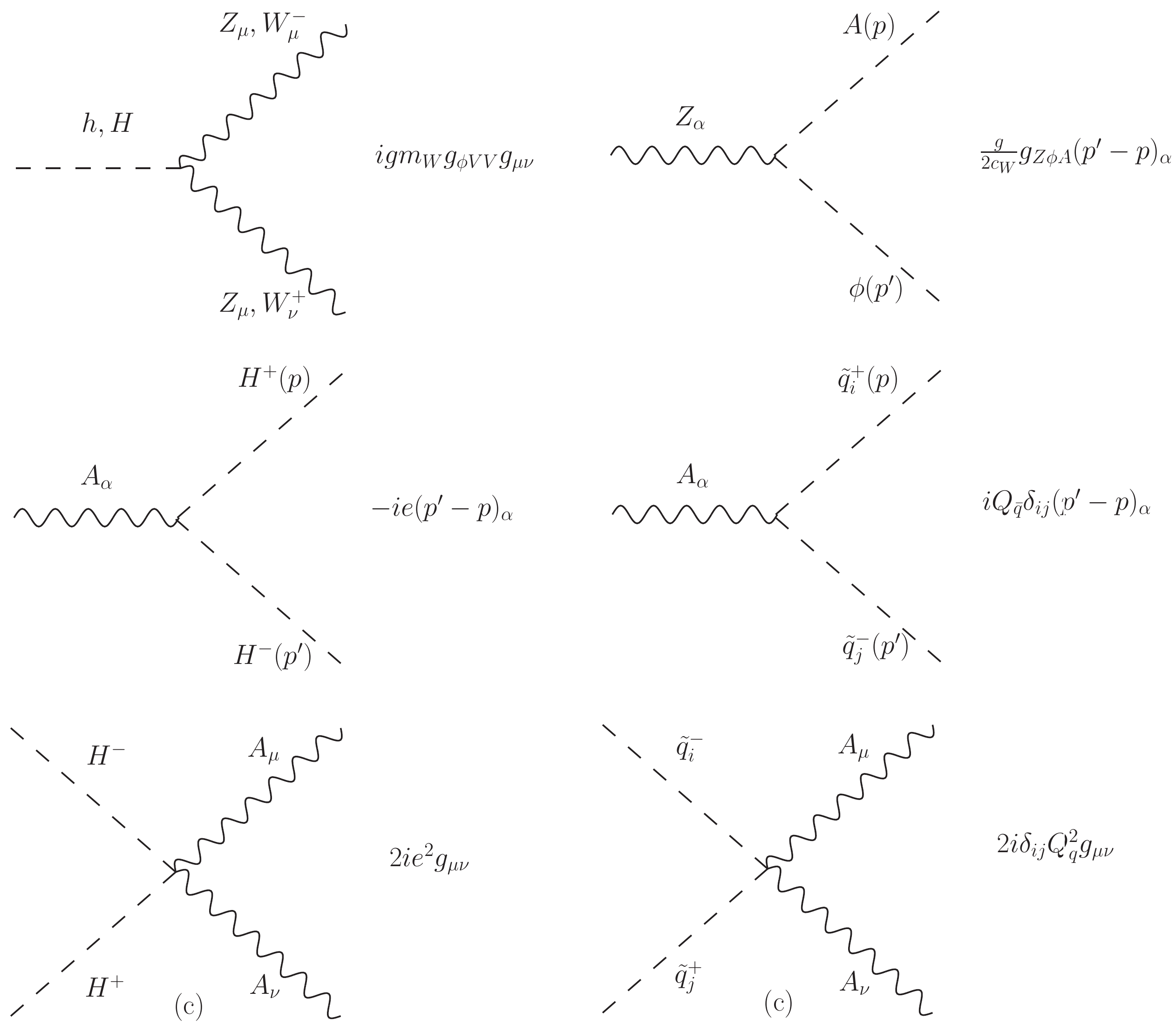}
\caption{Feynman rules for the couplings of scalar bosons and squarks to the gauge bosons in the MSSM. In the top-left plot $V=Z,W$ and in the top-right diagram $\phi=h,H$. All the four-momenta are incoming, and  the corresponding coupling constants are shown in Table \ref{CouplingConstants2}. }
\label{FRscalarboson}
\end{figure}

In Fig. \ref{FRscalartofermions} we present the Feynman rules for the couplings of the neutral Higgs bosons to fermions and  charginos, whereas the ones for  the couplings of scalar  bosons and squarks to gauge bosons, which are obtained by expanding the covariant derivative in terms of the physical fields, are shown in Fig. \ref{FRscalarboson}. Note that the $AVV$ ($V=W,Z$) and $HhZ$ couplings are absent due to $CP$ conservation.

\begin{table}[!hbt]
\begin{center}
\caption{Constants for the couplings of  scalar bosons to fermions and gauge bosons as described in Figs. \ref{FRscalartofermions} and \ref{FRscalarboson} along with Eq. \eqref{ScalarTrilinear}. Also $g_{\phi ZZ}=g_{\phi WW}/c_W^2$.}
\label{CouplingConstants2}
\begin{tabular}[t]{c c c c c c }
\hline
\hline
$\phi$  &$g_{\phi uu}$ 	 &$g_{\phi dd}$ ($g_{\phi ll}$)
&$g_{\phi WW}$ &$g_{\phi ZA}$ &$g_{\phi H^-H^+}$\\
\hline \hline
$h$ &$s_{\beta-\alpha}+\dfrac{c_{\beta-\alpha}}{t_\beta}$  &
$s_{\beta-\alpha}-t_\beta c_{\beta-\alpha}$ &
$s_{\beta-\alpha}$ &$c_{\beta-\alpha}$ &$\left(c_Wc_{\beta-\alpha}-\frac{1}{2c_W}c_{2\beta} c_{\beta+\alpha}\right)$
\\
$ H$ &$-\left(c_{\beta-\alpha}-\dfrac{s_{\beta-\alpha}}{ t_\beta}\right)$  &$-\left(c_{\beta-\alpha}+t_\beta s_{\beta-\alpha}\right)$ &$c_{\beta-\alpha}$ &$-s_{\beta-\alpha}$ &$ \left( c_W s_{\beta-\alpha}+\frac{1}{2c_W}c_{2\beta} s_{\beta+\alpha}\right) $\\
$A$  &$\dfrac{1}{t_\beta}$  &$t_\beta$ &$0$	&$0$ &$0$\\
\hline
\hline
\end{tabular}
\end{center}
\end{table}

\begin{table}[!hbt]
\begin{center}
\caption{Constants for the couplings of the neutral Higgs bosons to charginos. We have used the short-hand notation $s_a=\sin a$ and $ c_a=\cos a$.\label{charginocouplings}}
\begin{tabular}[t]{c c c}
\hline
\hline
$\phi$ &$g_{\phi\tilde{\chi}_i^+\tilde{\chi}_j^+}^S$ &$g_{\phi\tilde{\chi}_i^+\tilde{\chi}_j^+}^P$ \\
\hline
\hline
$A$ &$s_\beta(V_{i1}U_{j2}-V_{j1}^*U_{i2}^*)+c_\beta(V_{i2}U_{j1}-V_{j2}^*U_{i1}^*)$ &$s_\beta(V_{j1}^*U_{i2}^*+V_{i1} U_{j2})+c_\beta(V_{j2}^*U_{i1}^*+V_{i2}U_{j1})$\\
$H$ &$-ic_\alpha(V_{j1}^*U_{i2}^*+V_{i1}U_{j2})-is_\alpha(V_{j2}^*U_{i1}^*+V_{i2}U_{j1})$ &$ic_\alpha(V_{j1}^*U_{i2}^*-V_{i1}U_{j2})+is_\alpha(V_{j2}^*U_{i1}^*-V_{i2}U_{j1})$\\
$h$ &$is_\alpha(V_{i1}U_{j2}+V_{j1}^*U_{i2}^*)-ic_\alpha(V_{i2}U_{j1}+V_{j2}^*U_{i1}^*)$ &$is_\alpha(V_{i1}U_{j2}-V_{j1}^*U_{i2}^*)-ic_\alpha(V_{i2}U_{j1}-V_{j2}^*U_{i1}^*)$\\
\hline\hline
\end{tabular}
\end{center}
\end{table}

Finally, as for the trilinear  couplings involving scalar bosons and  sfermions, which arise from the scalar potential, the corresponding Feynman rules are
\begin{align}\label{ScalarTrilinear}
  \phi H^-H^+ &: -i \frac{g m_Z}{2c_W} g_{\phi H^-H^+}, \\
  \phi \bar{q}_i^-  {q}_j^+&: g_{\phi \tilde q_i \tilde q_j},
\end{align}
where  $\phi=h,H$ and the coupling constants $g_{\phi H^-H^+}$ are shown in Table \ref{CouplingConstants2}. In the notation of the third generation sfermions, the  constants $g_{\phi\tilde{q}_i\tilde{q}_j}$ read

\begin{equation}
\label{Cphiqiqj}
g_{\phi\tilde{q}_i\tilde{q}_j}=\frac{g_2}{m_W}\sum_{k,l=1}^2 (R^q)_{ik}^T C_{\phi\tilde{q}\tilde{q}^\prime}(R^{q^\prime})_{lj},
\end{equation}
with the matrices $C_{\phi\tilde{q}\tilde{q}^\prime}$ summarizing the couplings of the Higgs bosons to the squark  eigenstates.
In particular, the $H\tilde{t}\tilde{t}$ and $h\tilde{q}\tilde{q}$ coupling constants are
\begin{align}
g^{h\tilde{t}_1\tilde{t}_1}_{h\tilde{t}_2\tilde{t}_2}&=m_t^2 \mp \frac{m_t X_t\sin(2\theta)}{2} \pm  \frac{1}{12}m_Z^2\cos(2\beta)\bigl[(8s_W^2-3)\cos(2\theta)\pm3\bigr],\\
g^{H\tilde{t}_1\tilde{t}_1}_{H\tilde{t}_2\tilde{t}_2}&=\frac{m_t}{2t_\beta}\bigl[\pm\sin(2\theta)(A_t+\mu t_\beta)\mp2m_t)\bigr] \pm \frac{m_Z^2\sin(2\beta)}{12}\bigl[(8s_W^2-3)\cos(2\theta) \mp 3\bigr],
\end{align}
whereas the couplings of the $CP$-odd scalar boson to sfermions are

\begin{align}\label{Asqsqcoupling}
g_{A\tilde{t_1}\tilde{t_1}}&=g_{A\tilde{t_2}\tilde{t_2}}=0,\\
g_{A\tilde{t_2}\tilde{t_1}}&=-g_{A\tilde{t_1}\tilde{t_2}}=\frac{1}{2t_\beta^2}m_t(\mu+t_\beta(\mu t_\beta+X_t)).
\end{align}

\section{$A\to Z\gamma\gamma$ and $\phi\to Z\gamma\gamma$ decay amplitudes} \label{FFactors}

We now present the form factors $\mathcal{F}_i (i=1\ldots 7)$ of Eq. \eqref{amplitudeA} in terms of Passarino-Veltman scalar function. The contribution of SM charged fermions is identical to that arising in type-II THDMs, which was already presented in Ref. \cite{Sanchez-Velez:2018xdj}. We thus content ourselves with  presenting  the contributions of charginos and sfermions. The latter only contribute to reducible diagrams.

We first make the following definitions

\begin{equation}\label{SF}
\begin{split}
C_1(p^2) &=C_0(0, 0, p^2, m_{\tilde{\chi}_i^+}^2, m_{\tilde{\chi}_i^+}^2, m_{\tilde{\chi}_i^+}^2),\\
C_2(p^2)&=C_0(0,p^2, m_A^2 , m_{\tilde{\chi}_i^+}^2, m_{\tilde{\chi}_i^+}^2, m_{\tilde{\chi}_i^+}^2),\\
C_3(p^2)&=C_0(m_Z^2, 0,p^2, m_{\tilde{\chi}_i^+}^2, m_{\tilde{\chi}_i^+}^2, m_{\tilde{\chi}_i^+}^2),\\
C_4(p^2) &= C_0(m_Z^2, p^2, m_{\tilde{\chi}_i^+}^2, m_{\tilde{\chi}_i^+}^2, m_{\tilde{\chi}_i^+}^2),\\
D_1(p^2) &= D_0(m_Z^2, 0, 0, m_A^2 , p^2, s, m_{\tilde{\chi}_i^+}^2, m_{\tilde{\chi}_i^+}^2, m_{\tilde{\chi}_i^+}^2,m_{\tilde{\chi}_i^+}^2),\\
D_2(p^2) &= D_0(m_Z^2, 0, m_A^2, 0,s_1, p^2,  m_{\tilde{\chi}_i^+}^2, m_{\tilde{\chi}_i^+}^2, m_{\tilde{\chi}_i^+}^2,m_{\tilde{\chi}_i^+}^2).
\end{split}
\end{equation}
We also use the already  defined kinematic variables $s$, $s_1$, and $s_2$ [Eqs. (\ref{svariable})-(\ref{s2variable})] and  introduce the auxiliary variables $\Delta_{ij}$ and $X$ defined as

\begin{align}
\Delta_{ij}&=s_i-m_j\;\;\mbox{with}\;\; i=1,2\;\mbox{and} \; j=A,Z,\\
X_{\phi}&=s_1 s_2 -(m_Zm_\phi)^2.
\end{align}

\subsection{$A\to Z\gamma\gamma$ decay form factors}

The box diagrams give the following contributions to the form factors

\begin{equation}
\mathcal{F}_{i}^{\rm Box}=\sum_{\tilde{\chi}^+_i} \frac{16g^{\tilde{\chi}^+}_A g_{A\tilde{\chi}_i^+\tilde{\chi}_i^+} g^2 \alpha m_{\tilde{\chi}_i^+}^2}{\sqrt{2} c_W s X_A ^2} f_i^{\rm Box},
\end{equation}
with
\begin{align}\label{FFB1}
f_{1}^{\rm Box}&= s\Bigl[X_A \Delta_{2Z}+sm_Z^2(s_1+s_2)\Bigr]C_1(s)+{\frac{\Delta_{2A}}{ s}}\Delta_{2Z}\Bigl[s_2 \Delta_{1Z}^2-m_Z^2\Delta_{2A}^2\Bigr]C_2(s_2)+{\frac{\Delta_{1A}}{ s}}\Bigl[s_1 (s_1+s)\Delta_{2Z}^2\nonumber\\
&+[m_A^2(2s_1+s)(s_1-\Delta_{2Z})+(m_Z\Delta_{1A})^2-(s_1 s+(s_1+s)^2)s_1]m_Z^2 \Bigr]C_2(s_1)-\Delta_{1Z}^2\Bigl[\Delta_{2Z}\Bigl(m_Z^2-{\frac{X_A}{ s}}\Bigr)\nonumber\\
&+sm_Z^2\Bigr]C_3(s_1)+\Delta_{2A}\Delta_{2Z}^2\Bigl[m_Z^2-{\frac{X_A}{ s}}\Bigr]C_3(s_2)+sm_Z^2 \Bigl[(s_1-s_2)^2+X_A \Bigr]C_4(s)+{\frac{s}{ 2}}\Bigl[m_A^2m_Z^2(s_1(\Delta_{1Z}\nonumber\\
&+s_1-s_2)-2X_A )+s_1^2(2 m_Z^4-(2 s_1+s_2) m_Z^2+s_2^2)+\Bigl(m_Z^2-{\frac{X_A}{ s}}\Bigl)4m_{\tilde{\chi}_i^+}^2X_A \Bigr]D_1(s_1)-{\frac{s}{2}} \Bigl[s_1 s_2^3\nonumber\\
&-s_2m_Z^2(s_2(\Delta_{2Z}+s_1+3s)+X_A )-4Xm_{\tilde{\chi}_i^+}^2\Bigl(m_Z^2-{\frac{X_A}{ s}}\Bigr)\Bigr]D_1(s_2)+{\frac{1}{ 2}} \Bigl[s_2s^2(s_2-4 m_{\tilde{\chi}_i^+}^2)(m_Z^2+s_2)\nonumber\\
&+\Delta_{2A}\Delta_{2Z}\Bigl({\frac{2\Delta_{2A}^2}{ s}}\Delta_{2Z}^2-\Delta_{2A}(4 m_{\tilde{\chi}_i^+}^2-m_Z^2-5 s_2)\Delta_{2Z}+2(s_2-2m_{\tilde{\chi}_i^+}^2)(m_Z^2+2s_2)s\Bigr) \Bigr]D_2(s_2),
\end{align}

\begin{align}
f_{2}^{\rm Box}&=\frac{X_A}{2} \Bigl(2\Delta_{2A}\Bigl(sC_1(s) +\Delta_{1A}C_2(s_1)+{\frac{\Delta_{2Z}}{ 2}}C_3(s_2)\Bigr)+ sm_A^2\Delta_{1Z}D_1(s_1)\nonumber\\
&+\Delta_{1Z} X_A D_2(s_2)- s\Big[X_A +s_2\Delta_{2A}\Big]D_1(s_2)\Bigr),
\end{align}
and
\begin{align}\label{FFB3}
f_{3}^{\rm Box}&=2sm_A^2\Big[X_A +s(s_1+s_2)\Big]C_1(s)+2ss_1 \Delta_{1A}C_2(s_1) -2\Delta_{2A}^2\Delta_{2Z}C_2(s_2)\nonumber\\
&-2\Delta_{1Z}^2\Delta_{1A}C_3(s_1)+2ss_2\Delta_{2Z}C_3(s_2)-2s\Bigl[X_A +(s_1-s_2)^2\Bigr]C_4 (s)-X_A \Bigl[X_A+4s m_{\tilde{\chi}_i^+}^2\Bigr]D_2(s_2)\nonumber\\
&-s\Big[s_1 (X_A +2s_1s)+4X_A  m_{\tilde{\chi}_i^+}^2\Big]D_1 (s_1)- s\Big[s_2(X_A +2s_2s)+4 X_A m_{\tilde{\chi}_i^+}^2
 \Bigr]D_1(s_2),
 \end{align}

It is evident that  box diagram  amplitudes are free of ultraviolet divergences as they are free of two-point Passarino-Veltman scalar functions.

As for the reducible diagrams,  they  only contribute the form factor   $\mathcal{F}_1$, which  is defined in Eq.\eqref{F1RD} and receive the following  partial contributions from fermions $f$, the $W$ gauge boson, the charged scalar boson $H^\pm$,  charginos $\tilde\chi^+_i$ and squarks $\tilde q_i$:

\begin{equation}
\mathcal{F}_{1}^{\phi-\mathcal{X}}=\dfrac{2 \alpha g^2 g_{\phi ZA}} {\pi c_W s(m_\phi^2-s)}\left\{
\begin{array}{llll}\sum_f
\dfrac{g_{\phi \bar{f}f} m_f^2 Q_f^2N_c^f}{m_W}\Bigl[1+\Bigl(2m_f^2- \dfrac{s}{2}\Bigr)C_0(s,m_f^2)\Bigr]&&\mathcal{X}=f,\\
-\dfrac{g_{\phi WW}}{4 m_W}
\Bigl[ \dfrac{s}{2}+3m_W^2(1+(2m_W^2-s)C_0(s,m_W^2))\Bigr]&&\mathcal{X}=W,\\
\dfrac{g_{\phi H^\pm H^\pm}m_Z}{4 }
\Bigr[1+2m_{H^\pm}^2C_0(s,m_{H^\pm}^2)
\Bigr]&&\mathcal{X}=H^{\pm},\\
\sum_{\tilde{\chi}^+_i}-\dfrac{ig_{\phi\tilde{\chi}_i^+\tilde{\chi}_i^+}^S m_{\tilde{\chi}_i^+}}{\sqrt{2}}\Big[1+\left(2m_{\tilde{\chi}_i^+}^2-\frac{s}{2}\right)C_0(s,m_{\tilde{\chi}_i^+}^2) \Big]&&\mathcal{X}=\tilde\chi_i^+,\\
\sum_{\tilde{q}_i}-\dfrac{g_2
  g_{\phi \tilde{q}_i \tilde{q}_j} Q_{\bar q}^2}{4g m_W} \left[1+2
   m_{\tilde{q_i}}^2 C_0(s,m_{\tilde{q_i}}^2) \right]&&\mathcal{X}=\tilde q_i,
\end{array}\right.
\end{equation}
with $\phi=h,H$.

The three-point scalar function $C_0(s,m^2)$ can be written in terms of elementary functions as follows
\begin{equation}
C_0(s,m^2) =-\frac{2}{s} f\left(\frac{4m^2}{s}\right),
\end{equation}
with $f(x)$ being given by
\begin{equation}
\label{f(x)}
f(x)=\left\{
\begin{array}{cr}
\left[\arcsin\left(\frac{1}{\sqrt{x}}\right)\right]^2&x\ge1,\\
-\frac{1}{4}\left[\log\left(\frac{1+\sqrt{1-x}}{1-\sqrt{1-x}}\right)-i\pi\right]^2&x<1.
\end{array}
\right.
\end{equation}

\subsection{$H\to Z\gamma\gamma$ form factors}

The box diagram contributions to the form factors of Eq. \eqref{amplitudephi} are
\begin{equation}
\mathcal{G}_{i}^{\rm Box}=\sum_{\tilde{\chi}_i^+} \dfrac{16  g^2 \alpha g_A^{\tilde{\chi}_i^+}m_{\tilde{\chi}}g_{H\tilde{\chi}\tilde{\chi}} }{\sqrt{2} c_WX_{\phi}^2 }{g}_{i}^{\rm Box},
\end{equation}
with
\begin{equation}
\begin{split}
g_{1}^{\rm Box} &=\frac{X_{\phi}}{32}\Big(2 \Delta_{2{\phi} }C_2(s_2)-2 \Delta_{1{\phi} }C_2(s_1)+2
   (s_1-s_2) C_4(s)+\Delta_{2{\phi} }\Delta_{2Z}D_1(s_2)-\Delta_{1Z}\Delta_{1{\phi} }D_1(s_1)\Big),
\end{split}
\end{equation}

\begin{align}
g_{2}^{\rm Box} &=m_\phi^2\Biggl(-2\Big[X_{\phi} +s(s_1+s_2)\Big]C_1(s) +\dfrac{2}{s}\Delta_{1Z}\Delta_{1{\phi} }^2 C_2(s_1)+2s_2\Delta_{2{\phi} }C_2(s_2)-2\Delta_{1Z}sC_3(s_1)-\dfrac{2}{s}\Delta_{2Z}^2\Delta_{2{\phi} }C_3(s_2)\nonumber\\ &+2\Big[2X_{\phi} +(s_1-s_2)^2 \Big]C_4(s)+\Big[4m_{\tilde{\chi}_i^+}^2X_{\phi} +s_1(X_{\phi} +2s_1s) \Big]D_1(s_1)
-\Big[4m_{\tilde{\chi}_i^+}^2X_{\phi} +s_2(X_{\phi} +2s_2s) \Big]D_1(s_2)\nonumber \\
&+\dfrac{1}{s}\Big[ X_{\phi} (X_{\phi} +4sm_{\tilde{\chi}}^2) \Big]D_2(s_2)\Biggr),
\end{align}

\begin{align}
g_{3}^{\rm Box} &=\dfrac{1}{s^2}\Big(-s^2\Big[s_1s_2^2+m_Z^2(m_{\phi} ^2(m_Z^2+2s_2)+\Delta_{2Z}s_2) \Big]C_1(s)-\Delta_{1{\phi} }\Big[ m_Z^8-(3 s+2(s_1+s_2)) m_Z^6\nonumber\\
&+(2 s^2+3 s_1
   s+s_1^2+s_2^2+4 (s+s_1) s_2)
   m_Z^4-((s_1+s_2) s^2+s_2 (4
   s_1+s_2) s+2 s_1 s_2 (s_1+s_2))m_Z^2\nonumber\\
   &+s_1 (s+s_1) s_2^2\Big]C_2(s_1)+\Delta_{2{\phi} }\Delta_{2Z}\Big[ m_Z^2\Delta_{2{\phi} }^2-\Delta_{1Z}^2s_2 \Big]C_2(s_2)+\Delta_{2{\phi} }\Delta_{2Z}^2\Big[X_{\phi} -sm_Z^2 \Big]C_3(s_2)\nonumber\\
   &+\Delta_{1Z}^2\Big[sm_Z^2(s+2s_2-m_Z^2) -\Delta_{1Z}\Delta_{2Z}^2 \Big]C_3(s_1)-s^2m_Z^2\Big[ 2X_{\phi} +(s_1-s_2)^2  \Big]C_4(s)+\frac{1}{2} s \Big[4 X_\phi (X_\phi\nonumber\\
   &-sm_Z^2) m_{\tilde{\chi}}^2+s (-2
   m_Z^8+(4 s+3 s_1+4 s_2) m_Z^6-(2 s^2+3
   s_1 s+4 s_2 s+s_1^2+2 s_2^2+6 s_1 s_2) m_Z^4\nonumber\\
   &+s_1
   (3 s_2^2+3 s s_2+2 s_1 s_2-2 s s_1)
   m_Z^2-s_1^2 s_2^2)\Big] D_1(s_1)+
   \frac{1}{2} s \Big[4 X_\phi (X_\phi -sm_Z^2) m_{\tilde{\chi}}^2+s s_2
   (m_Z^4(s_2+m_\phi^2)\nonumber\\
   &-s_2 (3
   s+2 s_1+s_2) m_Z^2+s_1 s_2^2)\Big]D_1(s_2) +\frac{1}{2} X_\phi \Big[2 m_Z^8-(5 s+4
   (s_1+s_2)) m_Z^6+(3 s^2+4 m_{\tilde{\chi}_i^+}^2
   s+5 s_1 s\nonumber\\
   &+6 s_2 s+2 s_1^2+2 s_2^2+8 s_1 s_2)
   m_Z^4-(4 s (2 s+s_1+s_2) m_{\tilde{\chi}_i^+}^2+s_2
   (s^2+(6 s_1+s_2) s+4 s_1
   (s_1+s_2))) m_Z^2\nonumber\\
   &+s_1 s_2 (4
   s m_{\tilde{\chi}_i^+}^2+(s+2 s_1) s_2)\Big] D_2(s_2)\Big),
\end{align}
and
\begin{align}
g_4^{\rm Box} &=\dfrac{1}{8\Delta_{1Z}\Delta_{2Z}}\Big( 4m_Z^2X_{\phi}[ \Delta_{1Z}\Delta B(m_Z^2,s_2)+\Delta_{2Z}\Delta B(m_Z^2,s_1)]+4\Delta_{1Z}\Delta_{2Z}X_{\phi } \Delta B(s,m_A^2)\nonumber\\
&+2 s\Delta_{1Z}\Delta_{2Z}\Big[2X_\phi+m_Z^2(m_Z^2-s)-s_1(s_2
   +s_1) +s_2^2\Big]C_1(s)+\Delta_{1Z}\Delta_{2Z}
   (s+\Delta_{2Z})
   \Big[2s_1\Delta_{1Z}\nonumber\\
   &-X_\phi\Big]C_2(s_1)-\Delta_{1Z}\Delta_{2Z}(s+\Delta_{1Z})
   \Big[5X_\phi-2s_2(\Delta_{1Z}+s_1)+2s_2^2 \Big]C_2(s_2)
   -\Delta_{1Z}^2\Delta_{2Z}
   \Big[2s_1\Delta_{1Z}\nonumber\\
   &-X_\phi\Big]C_3(s_1)+\Delta_{1Z}\Delta_{2Z}^2
   \Big[5X_\phi-2s_2(\Delta_{1Z}+s_1)+2s_2^2 \Big]C_3(s_2)
   +2 \Delta_{1Z}  \Delta_{2Z}\Big[2m_Z^4(m_\phi^2+2(s_1+s_2))\nonumber\\
&-(5 s_1^2-3
   s_2^2+4 s (s_1-s_2))
   m_Z^2+(s_1-s_2) (s_1+s_2)^2\Big]C_4(s)+\Delta_{1Z}\Delta_{2Z}\Big[-2m_Z^6(m_Z^2+2m_\phi^2)\nonumber\\
&+(2 s^2+(5
   s_1+4 s_2) s+2 (s_1^2+4 s_2
   s_1+s_2^2)) m_Z^4-s_1 (s^2-(s_1-5
   s_2) s+4 s_2 (s_1+s_2))
   m_Z^2+s_1^2 (2 s_2^2\nonumber\\
&+s (s_2-2
   s_1))-4 m_{\tilde{\chi}_i^+}^2 \Delta_{1Z}
   X_\phi\Big]D_1(s_1)-\Delta_{1Z}\Delta_{2Z}\Big[-2m_Z^6(m_Z^2+2m_\phi^2)+(2
   (s+s_1)^2+2 s_2^2\nonumber\\
&+(9 s+8 s_1)
   s_2) m_Z^4-s_2 (5 s^2+3 (3
   s_1+s_2) s+4 s_1 (s_1+s_2))
   m_Z^2+s_2^2 (s_1 (s+2 s_1)+2 s
   s_2)+4 m_{\tilde{\chi}_i^+}^2 (m_Z^2\nonumber\\
&-2 s_1+s_2)
  X_\phi\Big]D_1(s_2)+4 m_{\tilde{\chi}_i^+}^2 \Delta_{1Z}\Delta_{2Z}
   (s_1-s_2)X_{\phi} D_2(s_2)
\Big),
\end{align}
with   $\Delta B(r_1^2,r_2^2)=B_0(r_1^2,m_{\tilde{\chi}_i^+}^2,m_{\tilde{\chi}_i^+}^2)-B_0(r_2^2,m_{\tilde{\chi}_i^+}^2,m_{\tilde{\chi}_i^+}^2)$.  It is thus evident that ultraviolet divergences cancel out.

As far a the reducible diagrams that induce the decay $\phi\to  Z\gamma\gamma$ ($\phi=h,H$) are concerned, they yield the following contributions to the  form factor ${\cal G}_3^{\rm RD}$ of Eq. \eqref{GRD}. The diagram mediated by the $CP$-odd scalar only receives contribution from charged fermions and charginos
\begin{equation}
\mathcal{G}_{3}^{A-\mathcal{X}}=\frac{g\alpha g_{\phi Z A}}{c_W\pi(s-m_A^2)}\left\{\begin{array}{lcr}\sum_f \dfrac{  g g_{A\bar{f}f}Q_f^2 m_f^2N_c^f}{ 2 m_Z}
C_0(s,m_f^2),&&\mathcal{X}=f\\
\sum_{\tilde{\chi}_i^+} m_{\tilde{\chi}_i^+}g^P_{A\tilde{\chi}_i^+\tilde{\chi}_i^+ }C_0(s,m_{\tilde{\chi}_i^+}^2)&&\mathcal{X}=\tilde{\chi}_i^,
\end{array}\right.
\end{equation}
whereas the diagram mediated by the $Z$ gauge boson yields
\begin{equation}
\mathcal{G}_{3}^{Z-f}=\dfrac{ g^2 \alpha g_{\phi ZZ} }{c_W \pi s m_Z }\sum_f Q_f^2 g^f_A m_f^2N_c^f
C_0(s,m_f^2).
\end{equation}

\section{ $\phi \to Z\gamma\gamma$ ($\phi=h,H,A$) squared average amplitude}
\label{squaredamplitude}

From the invariant amplitude for the $A \to Z\gamma\gamma$ decay given in  Eq. \eqref{amplitudeA}, we can readily obtain the square amplitude summed over photon and $Z$ polarizations, which is required for the calculation of the decay width \eqref{decaywidth}. The  result can be written as follows

\begin{align}
\label{M2ADecay}
|\overline{\mathcal{M}}(A\to Z\gamma\gamma)|^2&=\frac{m_A^6}{4}\Biggl(
\frac{\hat{s}^2\hat{\Delta}_{1Z}^2}{2 \mu _Z} \left| \mathcal{F}_1\right| ^2+
\frac{1}{2 \mu _Z}\lambda_3\left| \mathcal{F}_2\right| ^2+
\frac{1}{8 \mu _Z} \hat{\Delta}_{2Z}^2 \lambda_2\left| \mathcal{F}_3\right| ^2+
\frac{\hat{s}^2\lambda_1}{2 \mu _Z} {\rm Re}\left[ \mathcal{F}_1 \,\tilde{\mathcal{F}}_1^*\right]\nonumber\\&-
\frac{1}{2 \mu _Z} \hat{\Delta}_{1Z} \hat{\Delta}_{2Z}  \left( 2\hat{s} \mu _Z- \hat{\Delta}_{1Z}\hat{\Delta}_{2Z}\right){\rm Re}\left[ \mathcal{F}_2 \,\tilde{\mathcal{F}}_2^*\right]+
\frac{1}{8 \mu _Z}\lambda_1\lambda_2 {\rm Re}\left[ \mathcal{F}_3 \,\tilde{\mathcal{F}}_3^*\right]\nonumber\\&+
 \hat{s}^2 \hat{\Delta}_{1Z}{\rm Re}\left[ \mathcal{F}_1 \mathcal{F}_2^*\right]+
\frac{1}{2} \hat{s}^2\lambda_1 \lambda_2 {\rm Re}\left[ \mathcal{F}_1 \mathcal{F}_3^*\right]+
\frac{1}{2 \mu _Z}\hat{\Delta}_{1Z} \hat{\Delta}_{2Z}^2  \lambda_1\lambda_2{\rm Re}\left[ \mathcal{F}_2 \mathcal{F}_3^*\right]\nonumber\\&-
 \hat{s}^2 \hat{\Delta}_{1Z}{\rm Re}\left[ \mathcal{F}_1 \,\tilde{\mathcal{F}}_2^*\right]+
\frac{1}{2} \hat{s}^2\hat{\Delta}_{1Z}^2 {\rm Re}\left[ \mathcal{F}_1 \,\tilde{\mathcal{F}}_3^*\right]+
\frac{1}{2 \mu _Z}\hat{\Delta}_{1Z} {\rm Re}\left[ \mathcal{F}_2 \,\tilde{\mathcal{F}}_3^*\right]\Biggl)+\left(\hat{s}_1\leftrightarrow\hat{s}_2\right),
\end{align}
where $\hat{s}_i=s_i/m_A^2$,  $\hat{s}=s/m_A^2$, $\hat{\Delta}_{ij}={\Delta}_{ij}/m_A^2$,  $\tilde{\mathcal{F}}_i(s,s_1,s_2)=\mathcal{F}_i(s,s_2,s_1)$. Also

\begin{equation}
\label{lambda1}
\lambda_1=\mu _Z^2-\left(2\hat s+s_1+\hat s_2\right) \mu _Z+\hat s_1 \hat s_2,
\end{equation}

\begin{equation}
\label{lambda2}
\lambda_2=\left(\mu _Z^4-2 \left(\hat{s}+\hat{s}_1+\hat{s}_2\right) \mu _Z^3+\left(2 \hat{s}^2+2 \left(\hat{s}_1+\hat{s}_2\right) \hat{s}+\hat{s}_1^2+\hat{s}_2^2+4 \hat{s}_1 \hat{s}_2\right) \mu _Z^2-2 \hat{s}_1 \hat{s}_2 \left(\hat{s}+\hat{s}_1+\hat{s}_2\right) \mu _Z+\hat{s}_1^2 \hat{s}_2^2\right),
\end{equation}
and
\begin{equation}
\label{lambda3}\lambda_3=\hat{s}_1^2 \hat{\Delta}_{2Z}^2+2 \hat{s}_1 \left(\hat{s}-\hat{\Delta}_{2Z}\right) \mu _Z \hat{\Delta}_{2Z}+\mu _Z^2 \left(-2 \hat{s}^2+2 \mu _Z \hat{s}+\hat{s}_2^2+\mu _Z^2-2 \hat{s}_2 \left(\hat{s}+\mu _Z\right)\right).
\end{equation}

As for the $\phi \to Z\gamma\gamma$ ($\phi=h,H$) decay, we obtain from Eq. \eqref{amplitudephi} the following square amplitude summed over polarizations of the photons and the $Z$ gauge boson

\begin{align}
|\overline{\mathcal{M}}(\phi\to Z\gamma\gamma)|^2&=\frac{\hat s m_\phi^6}{2}\Bigl(
 \eta_2  \left| \mathcal{G}_1\right| ^2
-\frac{1}{4} \hat\Delta_{1Z}^2\eta_1 \left| \mathcal{G}_2\right| ^2
+\frac{\hat s \hat\Delta_{1Z}^2}{4 \mu _Z} \left| \mathcal{G}_3\right| ^2
+\frac{1}{4\hat s \mu _Z}\eta_3 \left| \mathcal{G}_4\right| ^2
-\eta_2  {\rm Re}\left[\mathcal{G}_1\tilde{\mathcal{G}}_1^*\right]
\nonumber\\&+ \hat\Delta _{2 Z}\eta_1 {\rm Re}\left[\mathcal{G}_1 \tilde{\mathcal{G}}_2^*\right]
-\hat s \hat\Delta _{2 Z}{\rm Re}\left[\mathcal{G}_1 \tilde{\mathcal{G}}_3^* \right]
- \eta_2  {\rm Re}\left[\mathcal{G}_1 \tilde{\mathcal{G}}_4^*\right]
+\frac{1}{8} \hat\Delta_{1Z} \hat\Delta _{2 Z} \eta_1 {\rm Re}\left[\mathcal{G}_2\tilde{\mathcal{G}}_2^*\right]
+\frac{1}{2} \hat s\eta_1 {\rm Re}\left[ \mathcal{G}_2 \tilde{\mathcal{G}}_3^*\right]
\nonumber\\&+ \hat\Delta_{1Z}
 \eta_1\frac{1}{2} \hat s {\rm Re}\left[\mathcal{G}_2 \tilde{\mathcal{G}}_4^*\right]
+\frac{1}{2 \mu _Z}  \left[\hat\Delta_{1Z} \hat\Delta _{2 Z}-2 \hat s \mu _Z\right]{\rm Re}\left[ \mathcal{G}_3 \tilde{\mathcal{G}}_3^*\right]
- \hat\Delta_{1Z}{\rm Re}\left[\mathcal{G}_3 \tilde{\mathcal{G}}_4^*\right]
\nonumber\\&
-\frac{1}{4 \hat s\mu _Z}\eta_3 {\rm Re}\left[\mathcal{G}_4 \tilde{\mathcal{G}}_4^*\right]
- \hat\Delta_{1Z}
 \eta_1{\rm Re}\left[\mathcal{G}_1 \mathcal{G}_2^*\right]+\hat s \hat\Delta_{1Z}{\rm Re}\left[\mathcal{G}_1 \mathcal{G}_3^*\right]
+\eta_2  {\rm Re}\left[\mathcal{G}_1 \mathcal{G}_4^*\right]
\nonumber\\&-\hat\Delta_{1Z}\eta_1 {\rm Re}\left[ \mathcal{G}_2 \mathcal{G}_4^*\right]
+\frac{1}{2} \hat s  \hat\Delta_{1Z} {\rm Re}\left[\mathcal{G}_3 \mathcal{G}_4^*\right]\Bigr)+\left(\hat{s}_1\leftrightarrow\hat{s}_2\right)
\end{align}
with $\tilde{\mathcal{G}}_i(s,s_1,s_2)=\mathcal{G}_i(s,s_2,s_1)$ and
\begin{equation}
\eta_1= \hat s\mu_Z-\hat\Delta_{1Z}\hat\Delta _{2 Z},
\end{equation}
\begin{equation}
\eta_2=2 \hat\Delta_{1Z} \hat\Delta _{2 Z}-\hat s \mu _Z,
\end{equation}

\begin{equation}
\eta_3=\left(\hat s_1^2 \hat\Delta _{2 Z}^2-2 \hat s_1 \left(\hat\Delta _{2 Z}-\hat s\right) \mu _Z \hat\Delta _{2 Z}+\mu _Z^2
  \left(\hat\Delta_{2Z} ^2-2\hat s( \hat\Delta_{2Z} +\hat s)\right)\right).
\end{equation}

\bibliography{biblio}

\begin{thebibliography}{41}
\expandafter\ifx\csname natexlab\endcsname\relax\def\natexlab#1{#1}\fi
\expandafter\ifx\csname bibnamefont\endcsname\relax
  \def\bibnamefont#1{#1}\fi
\expandafter\ifx\csname bibfnamefont\endcsname\relax
  \def\bibfnamefont#1{#1}\fi
\expandafter\ifx\csname citenamefont\endcsname\relax
  \def\citenamefont#1{#1}\fi
\expandafter\ifx\csname url\endcsname\relax
  \def\url#1{\texttt{#1}}\fi
\expandafter\ifx\csname urlprefix\endcsname\relax\def\urlprefix{URL }\fi
\providecommand{\bibinfo}[2]{#2}
\providecommand{\eprint}[2][]{\url{#2}}

\bibitem[{\citenamefont{Aad et~al.}(2012)}]{Aad:2012tfa}
\bibinfo{author}{\bibfnamefont{G.}~\bibnamefont{Aad}} \bibnamefont{et~al.}
  (\bibinfo{collaboration}{ATLAS}), \bibinfo{journal}{Phys. Lett.}
  \textbf{\bibinfo{volume}{B716}}, \bibinfo{pages}{1} (\bibinfo{year}{2012}),
  \eprint{1207.7214}.

\bibitem[{\citenamefont{Chatrchyan et~al.}(2012)}]{Chatrchyan:2012xdj}
\bibinfo{author}{\bibfnamefont{S.}~\bibnamefont{Chatrchyan}}
  \bibnamefont{et~al.} (\bibinfo{collaboration}{CMS}), \bibinfo{journal}{Phys.
  Lett.} \textbf{\bibinfo{volume}{B716}}, \bibinfo{pages}{30}
  (\bibinfo{year}{2012}), \eprint{1207.7235}.

\bibitem[{\citenamefont{Martin}(1997)}]{Martin:1997ns}
\bibinfo{author}{\bibfnamefont{S.~P.} \bibnamefont{Martin}}
  (\bibinfo{year}{1997}), \bibinfo{note}{[Adv. Ser. Direct. High Energy
  Phys.18,1(1998)]}, \eprint{hep-ph/9709356}.

\bibitem[{\citenamefont{Gunion and Haber}(1985)}]{Gunion85nuclearphysics}
\bibinfo{author}{\bibfnamefont{J.~F.} \bibnamefont{Gunion}} \bibnamefont{and}
  \bibinfo{author}{\bibfnamefont{H.~E.} \bibnamefont{Haber}},
  \emph{\bibinfo{title}{Nuclear physics b272 (1986) 1-76 north-holland,
  amsterdam higgs bosons in supersymmetric models (l)*}}
  (\bibinfo{year}{1985}).

\bibitem[{\citenamefont{Carena et~al.}(2012{\natexlab{a}})\citenamefont{Carena,
  Gori, Shah, and Wagner}}]{Carena2012}
\bibinfo{author}{\bibfnamefont{M.}~\bibnamefont{Carena}},
  \bibinfo{author}{\bibfnamefont{S.}~\bibnamefont{Gori}},
  \bibinfo{author}{\bibfnamefont{N.~R.} \bibnamefont{Shah}}, \bibnamefont{and}
  \bibinfo{author}{\bibfnamefont{C.~E.~M.} \bibnamefont{Wagner}},
  \bibinfo{journal}{Journal of High Energy Physics}
  \textbf{\bibinfo{volume}{2012}}, \bibinfo{pages}{14}
  (\bibinfo{year}{2012}{\natexlab{a}}), ISSN \bibinfo{issn}{1029-8479},
  \urlprefix\url{https://doi.org/10.1007/JHEP03(2012)014}.

\bibitem[{\citenamefont{Arbey et~al.}(2012)\citenamefont{Arbey, Battaglia,
  Djouadi, and Mahmoudi}}]{Arbey2012}
\bibinfo{author}{\bibfnamefont{A.}~\bibnamefont{Arbey}},
  \bibinfo{author}{\bibfnamefont{M.}~\bibnamefont{Battaglia}},
  \bibinfo{author}{\bibfnamefont{A.}~\bibnamefont{Djouadi}}, \bibnamefont{and}
  \bibinfo{author}{\bibfnamefont{F.}~\bibnamefont{Mahmoudi}},
  \bibinfo{journal}{Journal of High Energy Physics}
  \textbf{\bibinfo{volume}{2012}}, \bibinfo{pages}{107} (\bibinfo{year}{2012}),
  ISSN \bibinfo{issn}{1029-8479},
  \urlprefix\url{https://doi.org/10.1007/JHEP09(2012)107}.

\bibitem[{\citenamefont{Carena et~al.}(2012{\natexlab{b}})\citenamefont{Carena,
  Gori, Shah, Wagner, and Wang}}]{Carena:2012gp}
\bibinfo{author}{\bibfnamefont{M.}~\bibnamefont{Carena}},
  \bibinfo{author}{\bibfnamefont{S.}~\bibnamefont{Gori}},
  \bibinfo{author}{\bibfnamefont{N.~R.} \bibnamefont{Shah}},
  \bibinfo{author}{\bibfnamefont{C.~E.~M.} \bibnamefont{Wagner}},
  \bibnamefont{and} \bibinfo{author}{\bibfnamefont{L.-T.} \bibnamefont{Wang}},
  \bibinfo{journal}{JHEP} \textbf{\bibinfo{volume}{07}}, \bibinfo{pages}{175}
  (\bibinfo{year}{2012}{\natexlab{b}}), \eprint{1205.5842}.

\bibitem[{\citenamefont{Casas et~al.}(2013)\citenamefont{Casas, Moreno,
  Rolbiecki, and Zaldivar}}]{Casas:2013pta}
\bibinfo{author}{\bibfnamefont{J.~A.} \bibnamefont{Casas}},
  \bibinfo{author}{\bibfnamefont{J.~M.} \bibnamefont{Moreno}},
  \bibinfo{author}{\bibfnamefont{K.}~\bibnamefont{Rolbiecki}},
  \bibnamefont{and} \bibinfo{author}{\bibfnamefont{B.}~\bibnamefont{Zaldivar}},
  \bibinfo{journal}{JHEP} \textbf{\bibinfo{volume}{09}}, \bibinfo{pages}{099}
  (\bibinfo{year}{2013}), \eprint{1305.3274}.

\bibitem[{\citenamefont{Djouadi et~al.}(1998)\citenamefont{Djouadi, Driesen,
  Hollik, and Kraft}}]{Djouadi:1996yq}
\bibinfo{author}{\bibfnamefont{A.}~\bibnamefont{Djouadi}},
  \bibinfo{author}{\bibfnamefont{V.}~\bibnamefont{Driesen}},
  \bibinfo{author}{\bibfnamefont{W.}~\bibnamefont{Hollik}}, \bibnamefont{and}
  \bibinfo{author}{\bibfnamefont{A.}~\bibnamefont{Kraft}},
  \bibinfo{journal}{Eur. Phys. J.} \textbf{\bibinfo{volume}{C1}},
  \bibinfo{pages}{163} (\bibinfo{year}{1998}), \eprint{hep-ph/9701342}.

\bibitem[{\citenamefont{Abbasabadi and Repko}(2005)}]{Abbasabadi:2004wq}
\bibinfo{author}{\bibfnamefont{A.}~\bibnamefont{Abbasabadi}} \bibnamefont{and}
  \bibinfo{author}{\bibfnamefont{W.~W.} \bibnamefont{Repko}},
  \bibinfo{journal}{Phys. Rev.} \textbf{\bibinfo{volume}{D71}},
  \bibinfo{pages}{017304} (\bibinfo{year}{2005}), \eprint{hep-ph/0411152}.

\bibitem[{\citenamefont{Abbasabadi and Repko}(2008)}]{Abbasabadi2008}
\bibinfo{author}{\bibfnamefont{A.}~\bibnamefont{Abbasabadi}} \bibnamefont{and}
  \bibinfo{author}{\bibfnamefont{W.~W.} \bibnamefont{Repko}},
  \bibinfo{journal}{International Journal of Theoretical Physics}
  \textbf{\bibinfo{volume}{47}}, \bibinfo{pages}{1490} (\bibinfo{year}{2008}),
  ISSN \bibinfo{issn}{1572-9575},
  \urlprefix\url{https://doi.org/10.1007/s10773-007-9590-0}.

\bibitem[{\citenamefont{Sánchez-Vélez and
  Tavares-Velasco}(2018)}]{Sanchez-Velez:2018xdj}
\bibinfo{author}{\bibfnamefont{R.}~\bibnamefont{Sánchez-Vélez}}
  \bibnamefont{and}
  \bibinfo{author}{\bibfnamefont{G.}~\bibnamefont{Tavares-Velasco}},
  \bibinfo{journal}{Phys. Rev.} \textbf{\bibinfo{volume}{D97}},
  \bibinfo{pages}{095038} (\bibinfo{year}{2018}), \eprint{1802.01222}.

\bibitem[{\citenamefont{Gounaris et~al.}(2001)\citenamefont{Gounaris,
  Porfyriadis, and Renard}}]{Gounaris:2001rk}
\bibinfo{author}{\bibfnamefont{G.~J.} \bibnamefont{Gounaris}},
  \bibinfo{author}{\bibfnamefont{P.~I.} \bibnamefont{Porfyriadis}},
  \bibnamefont{and} \bibinfo{author}{\bibfnamefont{F.~M.}
  \bibnamefont{Renard}}, \bibinfo{journal}{Eur. Phys. J.}
  \textbf{\bibinfo{volume}{C20}}, \bibinfo{pages}{659} (\bibinfo{year}{2001}),
  \eprint{hep-ph/0103135}.

\bibitem[{\citenamefont{Djouadi et~al.}(2013)\citenamefont{Djouadi, Maiani,
  Moreau, Polosa, Quevillon, and Riquer}}]{Djouadi:2013uqa}
\bibinfo{author}{\bibfnamefont{A.}~\bibnamefont{Djouadi}},
  \bibinfo{author}{\bibfnamefont{L.}~\bibnamefont{Maiani}},
  \bibinfo{author}{\bibfnamefont{G.}~\bibnamefont{Moreau}},
  \bibinfo{author}{\bibfnamefont{A.}~\bibnamefont{Polosa}},
  \bibinfo{author}{\bibfnamefont{J.}~\bibnamefont{Quevillon}},
  \bibnamefont{and} \bibinfo{author}{\bibfnamefont{V.}~\bibnamefont{Riquer}},
  \bibinfo{journal}{Eur. Phys. J.} \textbf{\bibinfo{volume}{C73}},
  \bibinfo{pages}{2650} (\bibinfo{year}{2013}), \eprint{1307.5205}.

\bibitem[{\citenamefont{Patrignani et~al.}(2016)}]{Patrignani:2016xqp}
\bibinfo{author}{\bibfnamefont{C.}~\bibnamefont{Patrignani}}
  \bibnamefont{et~al.} (\bibinfo{collaboration}{Particle Data Group}),
  \bibinfo{journal}{Chin. Phys.} \textbf{\bibinfo{volume}{C40}},
  \bibinfo{pages}{100001} (\bibinfo{year}{2016}).

\bibitem[{\citenamefont{Csaki}(1996)}]{Csaki:1996ks}
\bibinfo{author}{\bibfnamefont{C.}~\bibnamefont{Csaki}}, \bibinfo{journal}{Mod.
  Phys. Lett.} \textbf{\bibinfo{volume}{A11}}, \bibinfo{pages}{599}
  (\bibinfo{year}{1996}), \eprint{hep-ph/9606414}.

\bibitem[{\citenamefont{Fayet}(2014)}]{Fayet:2014oua}
\bibinfo{author}{\bibfnamefont{P.}~\bibnamefont{Fayet}}, \bibinfo{journal}{Eur.
  Phys. J.} \textbf{\bibinfo{volume}{C74}}, \bibinfo{pages}{2837}
  (\bibinfo{year}{2014}).

\bibitem[{\citenamefont{Branco et~al.}(2012)\citenamefont{Branco, Ferreira,
  Lavoura, Rebelo, Sher, and Silva}}]{Branco:2011iw}
\bibinfo{author}{\bibfnamefont{G.~C.} \bibnamefont{Branco}},
  \bibinfo{author}{\bibfnamefont{P.~M.} \bibnamefont{Ferreira}},
  \bibinfo{author}{\bibfnamefont{L.}~\bibnamefont{Lavoura}},
  \bibinfo{author}{\bibfnamefont{M.~N.} \bibnamefont{Rebelo}},
  \bibinfo{author}{\bibfnamefont{M.}~\bibnamefont{Sher}}, \bibnamefont{and}
  \bibinfo{author}{\bibfnamefont{J.~P.} \bibnamefont{Silva}},
  \bibinfo{journal}{Phys. Rept.} \textbf{\bibinfo{volume}{516}},
  \bibinfo{pages}{1} (\bibinfo{year}{2012}), \eprint{1106.0034}.

\bibitem[{\citenamefont{Choi et~al.}(1999)\citenamefont{Choi, Djouadi, Dreiner,
  Kalinowski, and Zerwas}}]{Choi:1998ut}
\bibinfo{author}{\bibfnamefont{S.~Y.} \bibnamefont{Choi}},
  \bibinfo{author}{\bibfnamefont{A.}~\bibnamefont{Djouadi}},
  \bibinfo{author}{\bibfnamefont{H.~K.} \bibnamefont{Dreiner}},
  \bibinfo{author}{\bibfnamefont{J.}~\bibnamefont{Kalinowski}},
  \bibnamefont{and} \bibinfo{author}{\bibfnamefont{P.~M.}
  \bibnamefont{Zerwas}}, \bibinfo{journal}{Eur. Phys. J.}
  \textbf{\bibinfo{volume}{C7}}, \bibinfo{pages}{123} (\bibinfo{year}{1999}),
  \eprint{hep-ph/9806279}.

\bibitem[{\citenamefont{Djouadi}(2008)}]{Djouadi:2005gj}
\bibinfo{author}{\bibfnamefont{A.}~\bibnamefont{Djouadi}},
  \bibinfo{journal}{Phys. Rept.} \textbf{\bibinfo{volume}{459}},
  \bibinfo{pages}{1} (\bibinfo{year}{2008}), \eprint{hep-ph/0503173}.

\bibitem[{\citenamefont{Passarino and Veltman}(1979)}]{Passarino:1978jh}
\bibinfo{author}{\bibfnamefont{G.}~\bibnamefont{Passarino}} \bibnamefont{and}
  \bibinfo{author}{\bibfnamefont{M.~J.~G.} \bibnamefont{Veltman}},
  \bibinfo{journal}{Nucl. Phys.} \textbf{\bibinfo{volume}{B160}},
  \bibinfo{pages}{151} (\bibinfo{year}{1979}).

\bibitem[{\citenamefont{Mertig et~al.}(1991)\citenamefont{Mertig, Bohm, and
  Denner}}]{Mertig:1990an}
\bibinfo{author}{\bibfnamefont{R.}~\bibnamefont{Mertig}},
  \bibinfo{author}{\bibfnamefont{M.}~\bibnamefont{Bohm}}, \bibnamefont{and}
  \bibinfo{author}{\bibfnamefont{A.}~\bibnamefont{Denner}},
  \bibinfo{journal}{Comput. Phys. Commun.} \textbf{\bibinfo{volume}{64}},
  \bibinfo{pages}{345} (\bibinfo{year}{1991}).

\bibitem[{\citenamefont{Aad et~al.}(2015{\natexlab{a}})}]{Aad:2015zhl}
\bibinfo{author}{\bibfnamefont{G.}~\bibnamefont{Aad}} \bibnamefont{et~al.}
  (\bibinfo{collaboration}{ATLAS, CMS}), \bibinfo{journal}{Phys. Rev. Lett.}
  \textbf{\bibinfo{volume}{114}}, \bibinfo{pages}{191803}
  (\bibinfo{year}{2015}{\natexlab{a}}), \eprint{1503.07589}.

\bibitem[{\citenamefont{Aad et~al.}(2016{\natexlab{a}})}]{Khachatryan:2016vau}
\bibinfo{author}{\bibfnamefont{G.}~\bibnamefont{Aad}} \bibnamefont{et~al.}
  (\bibinfo{collaboration}{ATLAS, CMS}), \bibinfo{journal}{JHEP}
  \textbf{\bibinfo{volume}{08}}, \bibinfo{pages}{045}
  (\bibinfo{year}{2016}{\natexlab{a}}), \eprint{1606.02266}.

\bibitem[{\citenamefont{Aad et~al.}(2016{\natexlab{b}})}]{Aad:2015eda}
\bibinfo{author}{\bibfnamefont{G.}~\bibnamefont{Aad}} \bibnamefont{et~al.}
  (\bibinfo{collaboration}{ATLAS}), \bibinfo{journal}{Phys. Rev.}
  \textbf{\bibinfo{volume}{D93}}, \bibinfo{pages}{052002}
  (\bibinfo{year}{2016}{\natexlab{b}}), \eprint{1509.07152}.

\bibitem[{\citenamefont{Aad et~al.}(2014)}]{Aad:2014yka}
\bibinfo{author}{\bibfnamefont{G.}~\bibnamefont{Aad}} \bibnamefont{et~al.}
  (\bibinfo{collaboration}{ATLAS}), \bibinfo{journal}{JHEP}
  \textbf{\bibinfo{volume}{10}}, \bibinfo{pages}{096} (\bibinfo{year}{2014}),
  \eprint{1407.0350}.

\bibitem[{\citenamefont{Aad et~al.}(2015{\natexlab{b}})}]{Aad:2015jqa}
\bibinfo{author}{\bibfnamefont{G.}~\bibnamefont{Aad}} \bibnamefont{et~al.}
  (\bibinfo{collaboration}{ATLAS}), \bibinfo{journal}{Eur. Phys. J.}
  \textbf{\bibinfo{volume}{C75}}, \bibinfo{pages}{208}
  (\bibinfo{year}{2015}{\natexlab{b}}), \eprint{1501.07110}.

\bibitem[{\citenamefont{Aaboud et~al.}(2017)}]{Aaboud:2017yyg}
\bibinfo{author}{\bibfnamefont{M.}~\bibnamefont{Aaboud}} \bibnamefont{et~al.}
  (\bibinfo{collaboration}{ATLAS}), \bibinfo{journal}{Phys. Lett.}
  \textbf{\bibinfo{volume}{B775}}, \bibinfo{pages}{105} (\bibinfo{year}{2017}),
  \eprint{1707.04147}.

\bibitem[{\citenamefont{Aad et~al.}(2016{\natexlab{c}})}]{Aad:2015agg}
\bibinfo{author}{\bibfnamefont{G.}~\bibnamefont{Aad}} \bibnamefont{et~al.}
  (\bibinfo{collaboration}{ATLAS}), \bibinfo{journal}{JHEP}
  \textbf{\bibinfo{volume}{01}}, \bibinfo{pages}{032}
  (\bibinfo{year}{2016}{\natexlab{c}}), \eprint{1509.00389}.

\bibitem[{\citenamefont{Aad et~al.}(2016{\natexlab{d}})}]{Aad:2015kna}
\bibinfo{author}{\bibfnamefont{G.}~\bibnamefont{Aad}} \bibnamefont{et~al.}
  (\bibinfo{collaboration}{ATLAS}), \bibinfo{journal}{Eur. Phys. J.}
  \textbf{\bibinfo{volume}{C76}}, \bibinfo{pages}{45}
  (\bibinfo{year}{2016}{\natexlab{d}}), \eprint{1507.05930}.

\bibitem[{\citenamefont{Aaboud et~al.}(2018{\natexlab{a}})}]{Aaboud:2018ftw}
\bibinfo{author}{\bibfnamefont{M.}~\bibnamefont{Aaboud}} \bibnamefont{et~al.}
  (\bibinfo{collaboration}{ATLAS}) (\bibinfo{year}{2018}{\natexlab{a}}),
  \eprint{1807.04873}.

\bibitem[{\citenamefont{Aad et~al.}(2015{\natexlab{c}})}]{Aad:2015wra}
\bibinfo{author}{\bibfnamefont{G.}~\bibnamefont{Aad}} \bibnamefont{et~al.}
  (\bibinfo{collaboration}{ATLAS}), \bibinfo{journal}{Phys. Lett.}
  \textbf{\bibinfo{volume}{B744}}, \bibinfo{pages}{163}
  (\bibinfo{year}{2015}{\natexlab{c}}), \eprint{1502.04478}.

\bibitem[{\citenamefont{Khachatryan et~al.}(2014)}]{Khachatryan:2014wca}
\bibinfo{author}{\bibfnamefont{V.}~\bibnamefont{Khachatryan}}
  \bibnamefont{et~al.} (\bibinfo{collaboration}{CMS}), \bibinfo{journal}{JHEP}
  \textbf{\bibinfo{volume}{10}}, \bibinfo{pages}{160} (\bibinfo{year}{2014}),
  \eprint{1408.3316}.

\bibitem[{\citenamefont{Aaboud et~al.}(2018{\natexlab{b}})}]{Aaboud:2017sjh}
\bibinfo{author}{\bibfnamefont{M.}~\bibnamefont{Aaboud}} \bibnamefont{et~al.}
  (\bibinfo{collaboration}{ATLAS}), \bibinfo{journal}{JHEP}
  \textbf{\bibinfo{volume}{01}}, \bibinfo{pages}{055}
  (\bibinfo{year}{2018}{\natexlab{b}}), \eprint{1709.07242}.

\bibitem[{\citenamefont{Sirunyan et~al.}(2018)}]{Sirunyan:2018zut}
\bibinfo{author}{\bibfnamefont{A.~M.} \bibnamefont{Sirunyan}}
  \bibnamefont{et~al.} (\bibinfo{collaboration}{CMS}) (\bibinfo{year}{2018}),
  \eprint{1803.06553}.

\bibitem[{\citenamefont{Aad et~al.}(2015{\natexlab{d}})}]{Aad:2015pla}
\bibinfo{author}{\bibfnamefont{G.}~\bibnamefont{Aad}} \bibnamefont{et~al.}
  (\bibinfo{collaboration}{ATLAS}), \bibinfo{journal}{JHEP}
  \textbf{\bibinfo{volume}{11}}, \bibinfo{pages}{206}
  (\bibinfo{year}{2015}{\natexlab{d}}), \eprint{1509.00672}.

\bibitem[{\citenamefont{Amhis et~al.}(2014)}]{Amhis:2014hma}
\bibinfo{author}{\bibfnamefont{Y.}~\bibnamefont{Amhis}} \bibnamefont{et~al.}
  (\bibinfo{collaboration}{Heavy Flavor Averaging Group (HFAG)})
  (\bibinfo{year}{2014}), \eprint{1412.7515}.

\bibitem[{\citenamefont{Bhattacherjee et~al.}(2015)\citenamefont{Bhattacherjee,
  Chakraborty, and Choudhury}}]{Bhattacherjee:2015sga}
\bibinfo{author}{\bibfnamefont{B.}~\bibnamefont{Bhattacherjee}},
  \bibinfo{author}{\bibfnamefont{A.}~\bibnamefont{Chakraborty}},
  \bibnamefont{and}
  \bibinfo{author}{\bibfnamefont{A.}~\bibnamefont{Choudhury}},
  \bibinfo{journal}{Phys. Rev.} \textbf{\bibinfo{volume}{D92}},
  \bibinfo{pages}{093007} (\bibinfo{year}{2015}), \eprint{1504.04308}.

\bibitem[{\citenamefont{Hahn and Perez-Victoria}(1999)}]{Hahn:1998yk}
\bibinfo{author}{\bibfnamefont{T.}~\bibnamefont{Hahn}} \bibnamefont{and}
  \bibinfo{author}{\bibfnamefont{M.}~\bibnamefont{Perez-Victoria}},
  \bibinfo{journal}{Comput. Phys. Commun.} \textbf{\bibinfo{volume}{118}},
  \bibinfo{pages}{153} (\bibinfo{year}{1999}), \eprint{hep-ph/9807565}.

\bibitem[{\citenamefont{van Oldenborgh and
  Vermaseren}(1990)}]{vanOldenborgh:1989wn}
\bibinfo{author}{\bibfnamefont{G.~J.} \bibnamefont{van Oldenborgh}}
  \bibnamefont{and} \bibinfo{author}{\bibfnamefont{J.~A.~M.}
  \bibnamefont{Vermaseren}}, \bibinfo{journal}{Z. Phys.}
  \textbf{\bibinfo{volume}{C46}}, \bibinfo{pages}{425} (\bibinfo{year}{1990}).

\bibitem[{\citenamefont{Barman et~al.}(2016)\citenamefont{Barman,
  Bhattacherjee, Chakraborty, and Choudhury}}]{Barman:2016kgt}
\bibinfo{author}{\bibfnamefont{R.~K.} \bibnamefont{Barman}},
  \bibinfo{author}{\bibfnamefont{B.}~\bibnamefont{Bhattacherjee}},
  \bibinfo{author}{\bibfnamefont{A.}~\bibnamefont{Chakraborty}},
  \bibnamefont{and}
  \bibinfo{author}{\bibfnamefont{A.}~\bibnamefont{Choudhury}},
  \bibinfo{journal}{Phys. Rev.} \textbf{\bibinfo{volume}{D94}},
  \bibinfo{pages}{075013} (\bibinfo{year}{2016}), \eprint{1607.00676}.

\end{thebibliography}
\end{document}